\documentclass{iopart}
\usepackage{graphicx}
\usepackage{url}
\graphicspath{{./LowGraphics/}}
\usepackage{bm}
\usepackage{amsgen,amssymb,amsfonts,amsbsy}
\usepackage{hyperref}
\newcommand{\sout}[1]{}
\usepackage[usenames]{color}

\newcommand{\be}{\begin{equation}}
\newcommand{\ee}{\end{equation}}
\newcommand{\beq}{\begin{equation}}
\newcommand{\eeq}{\end{equation}}
\newcommand{\bea}{\begin{eqnarray}}
\newcommand{\eea}{\end{eqnarray}}
\newcommand{\bdm}{\begin{displaymath}}
\newcommand{\edm}{\end{displaymath}}

\newcommand{\eqref}[1]{(\ref{#1})}

\begin{document}

\title{Searches for Cosmic-String Gravitational-Wave Bursts in Mock LISA Data}

\author{Michael I Cohen${}^1$, Curt Cutler${}^{1,2}$, and Michele Vallisneri${}^{1,2}$}

\address{${}^1$Theoretical Astrophysics, California Institute of
Technology, Pasadena, California 91125}
\address{${}^2$Jet Propulsion, 4800 Oak Grove Dr., Pasadena, CA 91109}

\date{\today}

\begin{abstract}
A network of observable, macroscopic cosmic (super-)strings may well have formed in the early Universe.  If so, the cusps that generically develop on cosmic-string loops emit bursts of gravitational radiation that could be detectable by gravitational-wave interferometers, such as the ground-based LIGO/Virgo detectors and the planned, space-based LISA detector.
Here we report on two versions of a  LISA-oriented string-burst search pipeline that we have developed and tested within the context of the Mock LISA Data Challenges. 
The two versions rely on the publicly available  MultiNest and PyMC software packages, respectively.  To reduce the effective dimensionality of the search space, our implementations use the {\emph F}-statistic to analytically maximize over the signal's amplitude and polarization, $\mathcal{A}$ and $\psi$, and use the FFT to search quickly over burst arrival times $t_C$.
The standard {\emph F}-statistic is essentially a frequentist statistic that maximizes the likelihood;  
we also demonstrate an approximate, Bayesian version of the {\emph F}-statistic that incorporates realistic priors on $\mathcal{A}$ and $\psi$.  We calculate how accurately LISA can expect to measure the physical parameters of string-burst sources, and compare to results based on the Fisher-matrix approximation.  To understand LISA's angular resolution for string-burst sources, we draw maps of the waveform fitting factor [maximized over $(\mathcal{A}, \psi, t_C$)] as a function of sky position;  these maps dramatically illustrate why (for LISA) inferring the correct sky location of the emitting string loop will often be practically impossible.
In addition, we identify and elucidate several symmetries that are imbedded in this search problem, and we derive the distribution of cut-off frequencies $f_{\rm max}$ for observable bursts.
\end{abstract}

\pacs{04.30.Tv, 04.30.Db, 04.80.Nn, 98.80.Cq, 11.27.+d}

\section{Introduction}
\label{s:Introduction}

There are several mechanisms by which an observable network of cosmic (super)strings could have formed in the early Universe.  Basically, string formation arises from the breaking of some U(1) symmetry (either global or local) as the Universe expands and cools.  In the 1980s and 1990s, interest was primarily in cosmic strings arising from grand unified theories~\cite{VilenkinShellard}, but in recent years several string-theory-inspired inflationary models have also been shown to populate the Universe with a network of cosmic-scale strings~\cite{SarangiTye,Polchinski04}.  
For instance, brane-inflation models can naturally lead to the breaking of U(1) symmetries at the end of inflation, leading to the formation of both long fundamental strings and $D(k+1)$-branes that wrap around $k$ compact dimensions and extend in one of Nature's three large spatial dimensions. These long strings can be stable on cosmological timescales (depending on the exact model) and could reasonably have string tensions in the range $10^{-12} \lesssim \mu \lesssim 10^{-6}$.  We refer the reader to~\cite{Polchinski04_review} for a nice review of the main physical ideas.

Simulations have shown that string networks rapidly approach an attractor: the distribution of straight strings and loops rapidly becomes independent of its initial conditions.  The network properties {\it do} depend on two basic parameters of the strings, the string tension $\mu$ and the string reconnection probability $p$.   The distribution of loop sizes at their birth should in principle be derivable from $\mu$ and $p$, but the huge range of scales makes this a very difficult problem to solve via simulations, and today the typical loop size at birth (as a fraction of the Hubble scale) is still uncertain by many orders of magnitude.  We refer the reader to Allen~\cite{Allen_review} for a brief, pedagogical introduction to string networks, and to Vilenkin and Shellard \cite{VilenkinShellard} for a more comprehensive review.

Once formed, string loops oscillate and therefore lose energy and shrink due to gravitational-wave (GW) emission.  The spectrum of this GW background radiation is calculated to be roughly flat over many orders of magnitude in frequency, including the frequency bands where current ground-based GW interferometers (like LIGO and Virgo) and planned space-based GW interferometers (like LISA) are sensitive. 
It is conventional to express the energy density $\rho_{GW}$ of GWs in terms of 
\be 
\Omega_\mathrm{GW}(f) \equiv \frac{1}{\rho_c}\frac{d  \rho_\mathrm{GW}}{ d \, {\rm ln}\, f} \, ,
\ee
where $\rho_c$ is the Universe's closure density.
The current limit on  $\Omega_\mathrm{GW}(f)$ from pulsar timing is $\Omega_\mathrm{GW}(f\sim 2.5\times 10^{-7} \,\mathrm{Hz}) \lesssim  4\times 10^{-8}$~\cite{Jenet2006}, and the limit from first-generation ground-based interferometers is $\Omega_\mathrm{GW}(f\sim 100\,\mathrm{Hz}) < 6.9\times 10^{-6}$ \cite{2009Nature}.
For comparison, the Advanced LIGO detectors should be capable of detecting a stochastic background with $\Omega_\mathrm{GW}(f\sim 40 \, \mathrm{Hz}) \gtrsim 10^{-9}$~\cite{2009Nature}, while LISA should be capable of detecting a string-generated background 
$\Omega_\mathrm{GW}(f\sim 10^{-4}\mbox{--}10^{-1.5} \, \mathrm{Hz}) \gtrsim  10^{-10}$~\cite{HoganBender2001}.
(For LISA, this threshold is set \emph{not} by detector noise, but instead by the background from short-period Galactic binaries.)

In addition to this broadband stochastic  background,  Damour and Vilenkin \cite{DamourVilenkin2000,DamourVilenkin2001} pointed out that the \emph{kinks} and \emph{cusps} that form on cosmic strings produce short GW bursts that could also be detectable for a large range of string parameters $\mu$ and $p$.  Kinks are discontinuities in the string's tangent direction, which arise when strings overlap and interconnect, while cusps are points on the string that become instantaneously accelerated to the speed of light.  
The portion of string near the cusp beams a burst of linearly polarized GWs in a narrow cone around the cusp's direction of motion.  Damour and Vilenkin showed that, for current and planned GW interferometers,  cusp bursts should  be significantly more detectable than kink bursts, so for the rest of this paper we focus on the former. 
GW bursts from string cusps have a universal shape $h(t) \propto |t - t_C|^{1/3}$, or equivalently $\tilde h(f) = \mathcal{A} |f|^{-4/3} e^{2\pi i f t_C}$.  (More precisely, for observers that are not exactly at the center of the radiation cone, $\tilde h(f)$ carries a cut-off frequency $f_\mathrm{max}$ which also smooths out $h(t)$ at $t=t_C$; see Sec.\ 2 below.)
 
 Searches for cosmic-string bursts in LIGO--Virgo data are already being carried out, though to date there have been no detections~\cite{FirstLIGO2009}.  However it is easy to see that the planned space-based GW detector LISA should
 be far more sensitive to string bursts than any current or planned ground-based instrument,  due to two factors.
 To understand the first, recall that the matched-filtering signal-to-noise ratio (SNR) for
 any burst is given by 
 \be
\mathrm{SNR}^2 \sim \int_0^{f_\mathrm{max} } \frac{f^2 |\tilde h(f)|^2 \, d(\log f)}{f \, S_h(f) }
 \ee
for any single detector with noise spectral density $S_h(f)$,
up to geometrical factors $\sim 1$.  
Thus, for bursts with $ |\tilde h(f)| \propto f^{-4/3}$, we have (roughly) $\mathrm{SNR} \propto f_b^{-1/3}/ [f_b S_h(f_b)]^{1/2}$, where $f_b$ is the frequency where the detector has its best
sensitivity.  The value of  $f_b^{-1/3}/ [f_b S_h(f_b)]^{1/2}$ is $\sim 10$ times higher for LISA than Advanced LIGO, largely due
to LISA's much lower sensitive frequency band.  The second factor arises from the fact, discussed in Sec.\ 2, that a burst's cut-off frequency $f_{\rm max}$ scales as $\alpha^{-1/3}$, where $\alpha$ is the angular separation between the beam direction
(which is along the instantaneous direction of the cusp's motion) and the observer's line of sight.  From this, we will show in Sec.\ 2.4
that the rate of bursts arriving at the detector, and satisfying  $f_{\rm max} > f_b$, scales as $f_b^{-2/3}$.  Hence, based on a uniform Euclidean distribution of sources, we can estimate that the distance to the \emph{closest} burst that enters a detector's band scales as $f_b^{-2/9}$.
This is also a factor $\sim 10$ higher for LISA than Advanced LIGO.  So we conclude that in any given year, the 
strongest burst detected by LISA will have an SNR a factor $\sim 100$ larger than the strongest burst detected by Advanced LIGO.
Clearly,  LISA's much lower frequency range is a major advantage for string-burst searches.

%
%

 While individual bursts are relatively featureless, as Polchinski~\cite{Polchinski04_review} emphasizes, many burst detections would give us an approximate spectrum $dN/d\rho = \alpha \rho^{\beta}$  (where $N$ is the number of detections and $\rho$ is their SNR), and the two measured parameters $\alpha$ and $\beta$ in principle determine the fundamental string parameters $\mu$ and $p$, at least for networks that are dominated by a single type of string.  (However we note that in the large region of parameter space for which the strongest
 observed bursts would be much closer than the Hubble distance, the exponent $\beta$ must be very close to $-4$, and so measuring $\beta$ may not be very constraining on the
 underlying string parameters; see Sec.\ 2.3.) 
Also, there are large regions of parameter space for which LISA would detect \emph{both} individual string bursts from cusps \emph{and} the broadband stochastic background from loop oscillations~\cite{Siemens_to_be_published}.
Clearly the measured energy density of the background in the LISA band would place one additional constraint on the string model.

Since the gravitational waveforms from cusps are both very simple and rather precisely known, it is natural to search for them using matched filtering.  As we explain in more detail in Sec.\ 2, for any set of string parameters, one can easily compute the ${\rm SNR}^2$, which is essentially a measure of how well the model waveform (i.e., template) matches the data.  Then, roughly speaking, finding the best-fit parameters is a matter of maximizing the ${\rm SNR}^2$ over the six-dimensional source-parameter space.   For three of the parameters (the signal's amplitude $\mathcal{A}$, polarization $\psi$, and arrival time $t_C$), this maximization can be performed almost trivially, using a combination of the {\emph F}-statistic and the FFT. For the remaining three parameters (the two angles giving the source's sky position, and the cut-off frequency $f_\mathrm{max}$), we made use of two publicly available optimization codes: PyMC \cite{PyMC}, a Python implementation of Markov Chain Monte Carlo integration, and MultiNest \cite{FerozHobson2007,FerozHobson2008}, a Fortran 90 implementation of a multimodal nested-sampling algorithm \cite{Skilling}.  Employing two different optimization algorithms allowed us to carry out useful cross-checks.  For high-SNR cases, we were able to recover Fisher-matrix error estimates, as expected.

We tested our searches using data sets from the recent third Mock LISA Data Challenge (MLDC)~\cite{MLDC3_2008,MLDC4_2009}.  Both our PyMC and MultiNest searches performed well in locating the global SNR maxima in parameter space, and  our best-fit SNRs were within $1\%$ of the true SNRs for all MLDC3 cases. The sources proved difficult to localize correctly on the sky, but, as we show in Sec.\ 3, that was due to near-degeneracies intrinsic to the problem, rather than to a failure of our searches.

Two other reports on LISA string-burst searches, also developed and tested in the context of MLDC 3, have appeared recently \cite{KeyCornish2008,Feroz_strings_2009}.  Our work differs p from those in several ways: First, we use the {\emph F}-statistic and FFT to improve search efficiency.  Second, we present an in-depth analysis of waveform overlap (maximized over $\mathcal{A}$, $\psi$, and $t_c$) as a function of sky position. This analysis clarifies why, for most LISA cusp-burst detections, the source's sky location is likely to be very poorly constrained by the data. Third, we analyze in detail some aspects of the problem that heretofore have not been carefully explored, including a suite of nearly exact symmetries (most of which were not previously noted), and the expected distribution of the maximum frequency in {\it observed}  cusp-bursts.

Other authors have recently focused on other possible kinds of GW signatures from cosmic strings:   DePies and Hogan~\cite{DePiesHogan} pointed out that for very small string tensions ($10^{-19} \lesssim \mu \lesssim 10^{-11}$),  GWs might be detected from the oscillations of individual nearby strings, thanks to the nearly periodic nature of loop oscillations, and to the gravitational clustering of string loops near our Galaxy.  Leblond and colleagues \cite{Leblond2009} showed how the breaking of metastable cosmic strings
could result in detectable GW signals.  In this paper, however, we restrict attention to searches for cusp-bursts.

The plan for the rest of this paper is as follows:  In Sec.~\ref{s:theory} we briefly review the general form of a GW burst emitted by  a cosmic-string cusp, as well as the associated signal registered by LISA.  We also review how to maximize SNR cheaply over the extrinsic parameters $\mathcal{A}$, $\psi$, and $t_C$, using the {\emph F}-statistic and the FFT (both standard tricks), and we introduce an approximate Bayesian version of the {\emph F}-statistic, which is only slightly harder to compute than the standard variety.  Finally, we digress slightly to discuss the expected distribution of $f_{\rm max}$ for observable sources.  In Sec.~\ref{s:Symmetries} we discuss the near-degeneracies in the space of burst signals (and therefore in source parameter space), which significantly impact one's ability to infer the true source parameters from a measurement: to wit, there is a discrete near-symmetry between sky locations that are reflections of each other across the plane of the LISA detector; in addition, a typical signal from a generic sky location can be mimicked to surprising accuracy by templates corresponding to a broad swath of very distant points on the sky, if the amplitude, polarization and arrival time  of the templates are adjusted suitably.
In Sec.~\ref{s:Methods} we give brief reviews of the MCMC and nested-sampling search concepts, and we describe the particular tunings of these methods that we found to be efficient for our GW burst searches.   In Sec.~\ref{s:Results} we describe the efficacy and accuracy of our searches in the MLDC data sets.  We summarize our results and conclusions in Sec.~\ref{s:Conclusions}.   
Throughout this paper we use units where $G=c=1$; all quantities are expressed in units of seconds (to some power).

\section{Theoretical background}
\label{s:theory}

\subsection{The gravitational waveform from cosmic-string bursts}
\label{ss:Waveform}
The GWs arriving at the detector from string-cusp bursts are fully characterized by six parameters: the source's sky location (given in the MLDCs as the ecliptic latitude $\beta$ and longitude $\lambda$), the burst's overall amplitude (at the detector) $\mathcal{A}$, the polarization $\psi$, the burst's time of arrival $t_C$, and the upper cut-off frequency $f_\mathrm{max}$.

If we fix the direction $\hat{k}$ of GW propagation (i.e., we fix $\beta$ and $\lambda$) and we let $e^+_{ij}$  and  $e^{\times}_{ij}$ be a pair of orthogonal polarization basis tensors for waves traveling along $\hat{k}$, the general burst waveform is expressed most simply in the Fourier domain as 
\be\label{2pol}
\tilde h_{ij}(f) = \big[A^1 e^+_{ij}  + A^2 e^{\times}_{ij}\big] \Lambda(f) e^{2\pi i f t_C}, 
\ee
where we adopt the MLDC approximation for $\Lambda(f)$,
\bea\label{lambda}
\Lambda(f) \equiv
 \left\{
\begin{array}{cl}
	f^{-\frac{4}{3}} &f<f_\mathrm{max},\\
	f^{-\frac{4}{3}} e^{1-f/f_\mathrm{max}} & f>f_\mathrm{max}.
\end{array} \right.
\eea
\noindent
In terms of these variables, $\mathcal{A}$ and $\psi$ are given by
\be
\mathcal{A}  = \sqrt{(A^1)^2 + (A^2)^2}, \quad \psi = \arctan\big(A^2/A^1\big),
\ee
and in order of magnitude,
\be\label{mag}
\mathcal{A} \sim \frac{\mu L^{2/3}}{D_L}, \quad f_{\rm max} \sim 2/(\alpha^3 L),
\ee
where $\mu$ is the string tension, $L$ is the characteristic length of the cosmic string, $D_L$ is the luminosity distance to the cusp, and $\alpha$ is the angle between the observer and the center of the beam, which points along the cusp's instantaneous velocity.\footnote{What Damour and Vilenkin actually show is that $|\tilde h(f)| \propto f^{-4/3}$ for $f \ll f_\mathrm{max}$, and that  $|\tilde h(f))|$ falls to zero exponentially for $f \gg f_\mathrm{max}$.
Equation \eqref{lambda} follows the signal model implemented in the LIGO Algorithm Library (LAL) to generate burst injections. This model is more precise than Damour and Vilenkin's description, though not necessarily very accurate.  For consistency, the MLDCs adopted the LAL model.}

\subsection{Maximization over the extrinsic parameters}
\label{ss:Maximization}
The SNR can be maximized analytically over the parameters $\mathcal{A}$ and $\psi$ using a version of the {\emph F}-statistic, while the FFT provides a highly efficient method to maximize SNR over $t_C$. Let us work out the details, beginning with the {\emph F}-statistic.
Consider the space of cusp-burst waveforms, and fix the parameters $\Theta \equiv (\beta,\lambda,t_C,f_\mathrm{max})$.
We shall build a statistic that is equal to the log-likelihood maximized over the vector space of all $(A^1,A^2)$. This statistic is a straightforward adaptation of the method employed in the (more complicated) cases of circular-orbit binaries \cite{CuFl94} and GW pulsars \cite{jks,CutlerSchutz2005}.

The LISA science data will consist of the time series of laser-noise--canceling TDI observables (\cite{Vallisneri2005}, and references therein); all the available information about GWs can be recovered from a basis of three such observables, such as $A$, $E$, and $T$ \cite{Prince2002,VCT2008} (these three are especially expedient since they have uncorrelated noises). Thus we represent the detector output as the vector $\mathbf{s} \equiv \big(s_A(t), s_E(t), s_T(t)\big)$, and we define the natural inner product on the vector space of all possible LISA signals (see, e.g., \cite{Cutler_1998}),
\be\label{inner}
{\left\langle \mathbf{u} \,|\, \mathbf{v}\right\rangle} \equiv 2  \int_{-\infty}^{\infty}{\frac{\tilde u_A(f)\, v^*_A(f)\,df}{S_{A}(f)}  + \textrm{(integrals for $E$ and $T$)} \,,}
\ee
where $S_{A}(f)$ is the single-sided noise spectral density for the observable $A$ (and similarly for $S_{E}(f)$ and $S_{T}(f)$) .
Assuming Gaussian noise, the log-probability of any noise realization $\mathbf{n}$ is then just $(-1/2){\left\langle \mathbf{n} \,|\, \mathbf{n}\right\rangle}$, and therefore the log-likelihood of the data $\mathbf{s}$ given the signal model $\mathbf{h}$ is $(-1/2){\left\langle \mathbf{s} - \mathbf{h} \,|\, \mathbf{s} - \mathbf{h}\right\rangle}$. 

Now, both polarization components of the burst produce a linear response in the three TDI observables,
\begin{eqnarray}
A^1\Lambda(f) e^{2\pi i f t_C} e^+_{ij}   &\rightarrow&  A^1 \bigg(F^+_A, F^+_E, F^+_T\bigg) \Lambda(f) e^{2\pi i f t_C} \equiv A^1 \mathbf{h}_1(t_C), \\ 
A^2\Lambda(f) e^{2\pi i f t_C} e^{\times}_{ij} & \rightarrow& A^2 \bigg(F^{\times}_A, F^{\times}_E, F^{\times}_T \bigg) \Lambda(f) e^{2\pi i f t_C} \equiv A^2 \mathbf{h}_2(t_C); \nonumber
\end{eqnarray}
here the $F^{+,\times}_{A,E,T}$ are linear time-delay operators that encode the LISA response to plane GWs (see \cite{Vallisneri2005,geotdi}, as well as the discussion in Sec.\ \ref{ss:SkyPositionReflectionSymmetry}). The time delays change continuously as the LISA constellation orbits the Sun, but in the limit of short-lived GWs, LISA can be considered stationary, and the delays fixed. Thus, the operators can be represented as frequency-dependent complex factors $F^{+,\times}_{A,E,T}(t_C,f)$, which are the analogs of antenna patterns for ground-based interferometers. For cosmic-string bursts, this approximation is justified by the fact that most of the SNR is accumulated over several thousand seconds, to be compared with the one-year timescale of the LISA motion. In our searches, however, we always compute the full LISA response in the time domain, using \textit{Synthetic LISA} \cite{Vallisneri2005}.

The best-fit values of $A^1$ and $A^2$ are those that minimize
\be 
{\left\langle \mathbf{s} - A^1 \mathbf{h}_1(t_C) - A^2 \mathbf{h}_2(t_C) \,|\, \mathbf{s} - A^1 \mathbf{h}_1(t_C) - A^2 \mathbf{h}_2(t_C) \right\rangle} \, .
\ee
It is easy to show that the optimized $A^i$ and the log-likelihood $\log L$ are given by
\bea
A^i = \big(\Gamma^{-1}\big)^{ij} {\left\langle \mathbf{h}_j(t_C) \,|\, \mathbf{s}\right\rangle},  \\
\log L = - \frac{1}{2} \, \big[ {\left\langle \mathbf{s} \,|\, \mathbf{s}\right\rangle} -\big(\Gamma^{-1}\big)^{ij}  {\left\langle \mathbf{h}_i(t_C) \,|\, \mathbf{s}\right\rangle}{\left\langle \mathbf{h}_j(t_C)  \,|\, \mathbf{s}\right\rangle}\big] +\mathrm{const.} \label{logL},
\eea
\noindent where the constant in Eq.\ \eqref{logL} is just the logarithm of a volume factor, and where
\be
\Gamma_{ij}(t_C) = {\left\langle \mathbf{h}_i(t_C) \,|\, \mathbf{h}_j(t_C)\right\rangle}.
\ee
For any given data $\mathbf{s}$, the term ${\left\langle \mathbf{s} \,|\, \mathbf{s}\right\rangle}$ is also a constant; the remaining piece of $\log L$, which depends on $h$, is known as the {\emph F}-statistic, and it is given by
\be\label{defF}
F \equiv \frac{1}{2} \big(\Gamma^{-1}\big)^{ij}  {\left\langle \mathbf{h}_i(t_C) \,|\, \mathbf{s}\right\rangle}{\left\langle \mathbf{h}_j(t_C)  \,|\, \mathbf{s}\right\rangle}.
\ee
In the limit of high SNR, $F \approx {\rm SNR}^2/2$, while in the absence of GWs the expectation value of {\emph F} is 1. (It is 2 for GW pulsars, but in that case the {\emph F}-statistic is maximized analytically over twice as many parameters.)

Using the FFT to maximize SNR over the time of arrival is also a standard technique \cite{Smith1987}. Here we merely review the implementation details for our case.
We arrive at the best-fit $t_C$ [for a given $(\beta,\lambda,f_\mathrm{max})$] by a simple, iterative scheme. We make an initial estimate $t_C^{(0)}$ (e.g., by an initial search step in which the source is assumed to be at the ecliptic North pole), and we compute ${\tilde h}^{(0)}_1(f)$  and ${\tilde h}^{(0)}_2(f)$ using the time-delay operators evaluated for that time.
Next, we calculate the overlap integrals ${\left\langle \mathbf{h}_i(t_C) \,|\, \mathbf{s}\right\rangle}$ at times $t_C = t_C^{(0)} + \Delta t$ by taking the inverse Fourier transform,
\begin{equation} \fl
{\left\langle \mathbf{h}_i(t_C) \,|\, {\bf s}\right\rangle}  = 2\int_{-\infty}^{\infty}\bigg[ \frac{{\tilde s}_A(f)\, h^{(0)}_i(f)^* }{S_{A}(f)}  + \big(A \leftrightarrow E\big) + \big(A \leftrightarrow T\big)  \,  \bigg] \,  e^{-2\pi i f \Delta t} \, df \, .
\end{equation}
Approximating $\Gamma_{ij}$ as the constant $\Gamma_{ij}(t^{(0)}_C)$, we have
\begin{equation} \fl
F(t^{(0)}_C + \Delta t) =  \frac{1}{2}\big(\Gamma^{-1}(t^{(0)}_C)\big)^{ij}  {\left\langle \mathbf{h}_i\big(t^{(0)}_{C} + \Delta t \big) \,|\, \mathbf{s}\right\rangle}{\left\langle \mathbf{h}_j\big(t^{(0)}_C + \Delta t\big) \,|\, {\bf s}\right\rangle}.
\end{equation}
Of course, the advantage of this approach is that we can use the FFT to obtain $F(t^{(0)}_C + n \delta t)$ cheaply for all integers $n$, where $\delta t$ is the sampling time. We can now find the value $n = n_b$ that maximizes {\emph F}, fit a parabola to the values of {\emph F} at the points $n_b-1$, $n_b$, and $n_b+1$, and locate $\Delta t_b$ at the maximum of the parabola. We then set $t^{(1)}_C \rightarrow  t^{(0)}_C + \Delta t_b$, replace $(\Gamma^{-1}(t^{(0)}_C)\big)^{ij}$ by $(\Gamma^{-1}(t^{(1)}_C)\big)^{ij}$, and iterate.
The reason we are iterating is that we need to account for the change in the time-delay operators over the time $\Delta t$; in practice, we always find that the original estimate  $t^{(0)}_C$ is within $\sim 500$ s of the true $t_C$ (see Sec.\ \ref{ss:MCMC}), and that a single iteration determines the best-fit $t_C$ to $\sim 0.01$ s. (That is, further iterations change $t_C$ by $\lesssim 0.01$ s.)

This completes our account of the maximization of log-likelihood over the parameters $(\mathcal{A}, \psi, t_C)$. The search over the remaining parameters $(\beta,\lambda,f_\mathrm{max})$, is discussed in Sec.\ \ref{ss:MCMC}.

\subsection{Bayesian version of the $F$-statistic}
\label{ss:BayesianFstat}
As emphasized above, the {\emph F}-statistic maximizes the log-likelihood over the parameters $\mathcal{A}$ and $\psi$. However, since we have \emph{prior} information on their distribution, it makes sense to use it to improve their estimation, as well as detection performance. 
As shown by Prix and Krishnan \cite{Prix_Krishnan_2009}, it is straightforward to construct a Bayesian version of {\emph F} (which we shall call $F_B$) that incorporates the prior knowledge.
The exact form of $F_B$ is somewhat unwieldy, but in this paper we show how to construct an approximate version that is only slightly harder to compute than the standard {\emph F}-statistic, and that is quite accurate for reasonably high SNR (i.e., for the cases of greatest interest).

Given the LISA data $\mathbf{s}$, let $P(\Theta,\mathcal{A},\psi|\mathbf{s})$ be the posterior probability of the source parameters [with  $\Theta \equiv (\beta,\lambda,t_C,f_\mathrm{rm})$]. As per Bayes' theorem,
\begin{equation}
P(\Theta,\mathcal{A},\psi|\mathbf{s}) \propto
P(\mathbf{s}|\Theta,\mathcal{A},\psi) P(\Theta,\mathcal{A},\psi),
\end{equation}
where the first factor on the right is the likelihood of measuring $\mathbf{s}$ given the parameters, and the second is the prior parameter distribution.
Given rotational invariance (no preferred source direction, no preferred polarization, and no preferred angle between our line of sight and the cusp velocity vector), and given the scaling of $f_\mathrm{max}$ with the observing angle $\alpha$ given in Eq.\ \eqref{mag} (which implies that the solid angle $\alpha \, d\alpha$ is $\propto f_\mathrm{max}^{-5/3}\, df_\mathrm{max}$), the prior {\it must} have the general form
\begin{eqnarray}
P(\Theta,\mathcal{A},\psi) \, d\Theta \, d\mathcal{A} \, d\psi &=&
(\sin \beta \, d\beta) \, d\lambda \, dt_C (f_\mathrm{max}^{-5/3} df_\mathrm{max}) \label{eq:prior} \\
&\times& (w(\mathcal{A}) d\mathcal{A}) \, d\psi, \nonumber
\end{eqnarray}
where $w(\mathcal{A})$ is a function of $\mathcal{A}$ that encodes cosmological information. For simplicity, in the rest of this paper we shall set $w(\mathcal{A}) = \mathcal{A}^{-4}$, as appropriate for a uniform distribution of strings in Euclidean space ($\mathcal{A} \propto r^{-1}$, where $r$ is the distance to the source, implies $r^2 dr \propto \mathcal{A}^{-4}\,d\mathcal{A}$).
This is a reasonable approximation for light strings ($\mu \lesssim 10^{-8}$), for which the strongest bursts that LISA observes would occur at $z < 1$.   It is  straightforward to modify the calculation below to treat any other form of $w(\mathcal{A})$.
The Bayesian version of the {\emph F}-statistic corresponds to \emph{integrating the posterior} $P(\Theta,\mathcal{A},\psi|\mathbf{s})$ over $\mathcal{A}$ and $\psi$, as opposed to \emph{maximizing the likelihood} for the regular {\emph F}-statistic.
Fixing the data $\mathbf{s}$ and the parameters $\Theta$, let $\mathbf{h}_b$ be the best-fit waveform with the $\mathcal{A}_b$ and $\psi_b$ that minimize ${\left\langle \mathbf{s} - \mathbf{h} \,|\, \mathbf{s} - \mathbf{h} \right\rangle}$. Defining $\Delta \mathbf{h} \equiv \mathbf{h}(\Theta,\mathcal{A}_b,\psi_b) - \mathbf{h}(\Theta,\mathcal{A},\psi) \equiv \mathbf{h}_b - \mathbf{h}$, we have
\bea
{\left\langle \mathbf{s} - \mathbf{h} \,|\, \mathbf{s} - \mathbf{h}\right\rangle} & \equiv & {\left\langle \mathbf{s} - \mathbf{h}_b + \Delta \mathbf{h} \,|\, \mathbf{s} - \mathbf{h}_b + \Delta \mathbf{h}\right\rangle} \nonumber\\
&=& {\left\langle \mathbf{s} - \mathbf{h}_b \,|\, \mathbf{s} - \mathbf{h}_b\right\rangle}  + {\left\langle \Delta \mathbf{h} \,|\, \Delta \mathbf{h}\right\rangle} \label{corr2}\\ 
& =& {\left\langle \mathbf{s} \,|\, \mathbf{s}\right\rangle} -  2 F + {\left\langle \Delta \mathbf{h} \,|\, \Delta \mathbf{h}\right\rangle}  \, ; \label{corr3}
\eea
here Eq.\ \eqref{corr2} holds because $\Delta \mathbf{h}$ lies in the $(A^1,A^2)$ vector subspace, to which $\mathbf{s} - \mathbf{h}_b$ is orthogonal thanks to the best-fit condition, and Eq.\ \eqref{corr3} follows from the very definition of {\emph F}. Thus, the Bayesian $F_B$ is defined by
\be
e^{F_B(\Theta)} = e^{F(\Theta)} \int e^{-\left\langle \Delta \mathbf{h} \,|\, \Delta \mathbf{h}\right\rangle /2} \mathcal{A}^{-4} \, d\mathcal{A} \, d\psi \, ,
\ee
\noindent
or equivalently
\be\label{fb2}
F_B(\Theta) = F(\Theta) - \log \left[
\int e^ {-\Gamma_{ij} \delta A^i \delta A^j / 2} \mathcal{A}^{-5} dA^1 dA^2
\right] \, ,
\ee
where we have changed variables from $(\mathcal{A},\psi)$ to $(A^1,A^2)$, defined $(A_b^1,A_b^2)$ to be the best-fit values of the amplitude parameters and $\delta A^i \equiv A^i - A_b^i$, used the definition of $\Gamma^{ij}$, and transformed volume elements using the standard identity $dA^1 dA^2 = \mathcal{A} \, d\mathcal{A} \, d\psi$. We shall now introduce an approximation that is appropriate in the limit of high SNRs, for which the exponential $e^{-\Gamma_{ij} \delta A^i \delta A^j / 2}$ becomes ever more peaked around $\delta A^i = 0$. We therefore expand $\mathcal{A}^{-5}$ around $\mathcal{A}_b$, discarding all terms higher than quadratic:
\be\label{expand}
\mathcal{A}^{-5} \rightarrow \mathcal{A}_b^{-5}  + \delta A^i\partial_i(\mathcal{A}^{-5})|_{\mathcal{A}_b}  + \frac{1}{2} \delta A^i \delta A^j \partial_i \partial_j
(\mathcal{A}^{-5})|_{\mathcal{A}_b }\, .
\ee
\noindent
Note that this approximation effectively regularizes the divergence of $P(\Theta,\mathcal{A},\psi \,|\,\mathbf{s})$ as $\mathcal{A} \rightarrow 0$, which arises from the $\mathcal{A}^{-4}$ factor in the integrand. This divergence is unphysical anyway; it originates in the assumption of an infinite Euclidean universe, and so it is basically another version of Olbers' paradox. If we had used a cosmologically sensible prior, such as one based on an FRW universe, there would have been no divergence in the first place.

Because of symmetry, the linear term (and indeed all odd terms) of Eq.\ \eqref{expand} brings no contribution to the Gaussian integral. Compared to the zeroth-order term, the contribution of the quadratic term is suppressed by $O(\mathrm{SNR})^{-2}$, and the contribution of the quartic piece by $O(\mathrm{SNR})^{-4}$, which justifies neglecting the latter. The remaining integral is trivial: defining
\be
\lambda_{ij} \equiv \frac{1}{2} \mathcal{A}_b^5 \, \partial_i \partial_j 
(\mathcal{A}^{-5})|_{\mathcal{A}_b} =
\frac{35}{2} \mathcal{A}_b^{-4} (A_b)_i (A_b)_j - \frac{5}{2} \mathcal{A}_b^{-2} \delta_{ij},
\ee
we have
\begin{equation} \fl
\mathcal{A}_b^{-5}  \int e^ {-\Gamma_{ij} \delta A^i \delta A^j/2} \big[ 1 + \lambda_{ij}\big]  d(\delta A^1) d(\delta A^2)  = 
2 \pi \mathcal{A}_b^{-5} (\det \Gamma)^{-1/2} [ 1 + \lambda_{ij} (\Gamma^{-1})^{ij}\big], 
\end{equation}
and therefore
\be
F_B =  F  - 5 \log \mathcal{A}_b - \frac{1}{2} \log \det \Gamma + \log \big[1 +  \lambda_{ij} (\Gamma^{-1})^{ij}\big] \, .
\label{eq:fb}
\ee
where we have ignored the constant $\log \pi$ term, which is irrelevant to searches.
Aesthetically, the reader may prefer to multiply the integral by a constant scale factor $s^3$, where $s$ is typical size for $\mathcal{A}$ and the $A^i$ (e.g., $10^{-21}$), and then work with rescaled versions of $\mathcal{A}$ , $A^i$, $\Gamma_{ij}$, and $\lambda_{ij}$, so that these are all within a few orders of magnitude of unity: $\bar \mathcal{A} \equiv \mathcal{A}/s$, $\bar A^i \equiv A^i/s$, $\bar \Gamma_{ij} \equiv s^2 \, \Gamma_{ij}$, and $\bar \lambda_{ij} \equiv s^2 \lambda_{ij}$. This leads to an equivalent representation of $F_B$, given by Eq.\ \eqref{eq:fb} after replacing all variables with their barred version.

The effect of the ``Bayesian correction'' terms in $F_B$ is to penalize fits that have relatively larger amplitude parameters $A^i$. This is precisely what we should expect: since the amplitudes scale as $1/r$, larger $A^i$ must come from strings that inhabit smaller volumes around the detector, which is a priori less likely. Note also that the terms involving $\Gamma_{ij}$ (or its inverse or determinant) incorporate the effects of the detector response, and therefore depend on sky location; for the same $A^i$, they penalize sky-locations for which the LISA response is relatively poorer.

Ironically, our Bayesian correction is not quite appropriate for the sources in MLDC data sets, which have SNRs drawn from a uniform distribution, so that farther sources are \emph{not} more likely that nearby ones, and sources from sky locations with a poor LISA response are equally likely to be detected. Thus, while our $F_B$ (or its analog with a better cosmological model) would be optimal for a real search, it does not minimize the expected parameter-estimation error for our MLDC entries.
 
\subsection{Distribution of $f_{\rm max}$ for detected bursts}
\label{ss:fmaxdistribution}
\begin{figure}
\centerline{\includegraphics[width=0.55\textwidth]{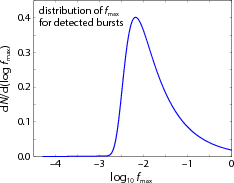}}
\caption{Expected distribution $dN/d(\log(f_\mathrm{max})$ of the maximum burst frequency $f_\mathrm{max}$ for the string bursts detectable by LISA.
\label{fig:fmax}}
\end{figure}

As an enlightening application of the distribution of burst parameters given in Eq.\ \eqref{eq:prior}, we estimate the distribution of the cut-off frequency $f_{\rm max}$ for the cosmic-string bursts that LISA would actually detect; i.e., for the bursts whose SNR is above some detection threshold $\rho_{th}$. We shall see that for most detections $f_{\rm max}$ is in-band and is $< 50$ mHz.
Since this subsection is something of a digression from the main flow of this paper, we are content with providing a sketch of the derivation.
 
The first step is to change variables from ${\mathcal A}$ to $\rho$, where $\rho$ is the SNR of the observation (the other five parameters remain the same). Clearly $\rho \propto {\mathcal A}$.  For simplicity, we estimate $\rho$ in the low-frequency approximation to the LISA response~\cite{Cutler_1998}.  In this approximation, the response functions factorize into a frequency-dependent term times an angle-dependent term, so we can write
\be
\rho = {\mathcal A}\, \eta(f_{\rm max}) \kappa(\beta, \lambda, \psi) \,
\ee
\noindent
where $\kappa$ is a known function of the angles $(\beta, \lambda, \psi)$ whose precise form is irrelevant, and 
\be
\eta(f_{\rm max}) \equiv  \bigg[\int_0^{f_\mathrm{max} } \frac{\Lambda^2(f) \, d f}{S_h(f) } \bigg]^{1/2}  \, ,
\ee
\noindent 
where $\Lambda(f)$ was defined in Eq.\ \eqref{lambda}, and $S_h(f)$ [unlike the $S_{A,E,T}(f)$ of Eq.\ \eqref{inner}] includes the frequency-dependent LISA response.  The Jacobian of the transformation is just $(\eta \kappa)^{-1}$.  Integrating the prior over all the angles, over the observation time, and over $\rho$ from the detection threshold $\rho_{th}$ up to $\infty$, we are left with the probability distribution of detectable bursts
\be
dN/d f_{\rm max} \propto   f^{-5/3}_{\rm max} \eta^3(f_{\rm max}) \, .
\ee
In Fig.\ \ref{fig:fmax} we plot the function $dN/d(\log f_{\rm max})$.  To evaluate $\eta$, we used the $S_h(f)$ fit given in Eqs.\ (26)--(31) of \cite{BarackCutler2004}, which includes confusion noise from unresolved white-dwarf binaries, and for simplicity we approximated $\Lambda(f)$ as $f^{-4/3} \Theta(f_{\rm max} -f)$, with  $\Theta(f_{\rm max} -f)$ the Heaviside function.  As $f_{\rm max}$ increases above $\sim 10$ mHz, $\eta$ remains nearly constant, so at these higher frequencies $dN/d(\log f_{\rm max})$ scales as $f^{-2/3}_{\rm max}$.  We find that the median value of $f_{\rm max}$ is $12\,$mHz, and that $\sim 2/3$ of detected string bursts will have $f_{\rm max} \in [5,50]$ mHz.

For this calculation we have assumed the ``uniform, Euclidean'' prior on the amplitude, $w({\mathcal A}) \propto {\mathcal A}^{-4}$; however it should be clear that the qualitative conclusion would remain the same even if most detected bursts were at cosmological distances.  Of course, the results for the case of ground-based detectors like LIGO and Virgo would be completely analogous: the median $f_{\rm max}$ for detected string bursts should be a factor $\sim 2\mbox{--}3$ higher than the frequency where $S_h(f)$ is at a minimum. 
Since for both ground-based and space-based GW detectors $f_{\rm max}$ will be in-band for most observed bursts, it seems worthwhile to devote more effort to determining the precise shape of $\tilde h(f)$ around $f_{\rm max}$ (instead of just patching together a power law with an exponential, as is currently done).

\section{Near-symmetries and overlap maps}
\label{s:Symmetries}

\subsection{Sky-position reflection across the LISA plane}
\label{ss:SkyPositionReflectionSymmetry}

There is a degeneracy in the LISA response to short-duration, linearly polarized GW sources that are located at sky positions related by a reflection across the LISA plane, as first noted in~\cite{KeyCornish2008}.  This degeneracy  becomes exact in the limit of infinitely short (and linearly polarized) GW signals. To understand how this degeneracy arises, we recall that the GW response of the laser-noise--canceling TDI observables can be written as~\cite{Vallisneri2005}
\begin{equation}
\mathrm{TDI}(t) = \sum_A c_A \, y_{(slr)_A}(t - \Delta_A),
\end{equation}
where the $y_{slr}(t)$ denote the one-way phase measurements along the six LISA laser links; the $slr$ triplet (a permutation of $123$) indexes the laser-\textit{s}ending spacecraft, the \textit{l}ink, and the \textit{r}eceiving spacecraft (see Fig.\ 3 of \cite{Vallisneri2005}); the $\Delta_A$ are time delays (sums of the inter-spacecraft times of flight), and $c_A = \pm 1$. Each phase measurement $y_{slr}$ registers plane GWs according to
\begin{equation}
y_{slr}(t) =
\frac{
\hat{n}_l(t) \cdot \bigl[ h\bigl(t_s - \hat{k} \cdot p_s(t_s)\bigr)
- h\bigl(t - \hat{k} \cdot p_r(t)\bigr) \bigr] \cdot \hat{n}_l(t)
}{
2\bigl(1 - \hat{k}\cdot\hat{n}_l(t)\bigr)
}.
\label{eq:responses}
\end{equation}
To parse this equation, it is useful to think about the effect of GWs on a single laser pulse received at spacecraft $r$ at time $t$: the unit vector $\hat{k}$ points along the direction of GW propagation; $h$ is the GW strain tensor at the solar system barycenter (SSB), which is transverse to $\hat{k}$; the $p_{s,r}(t)$ are the positions of the sending and receiving LISA spacecraft; the $\hat{n}_l(t) \propto p_r(t) - p_s(t_s)$ are the photon-propagation unit vectors; and the retarded time $t_s$ is determined by the light-propagation equation $t_s = t - |p_r(t) - p_s(t_s)|$. Thus, the GW strain tensor $h$ is projected onto $\hat{n}_l$ at the events $(t,p_r(t))$ and $(t_s,p_s(t_s))$ [the reception and emission of the pulse]. For plane GWs, the value of $h$ at those events is obtained by giving $h$ the appropriate retarded-time arguments $t - \hat{k} \cdot p_r(t)$ and $t_s - \hat{k} \cdot p_s(t_s)$.

Because the $p_i(t)$ evolve on the LISA orbital timescale of a year, LISA can be considered stationary with respect to signals of much shorter duration. In that case, the three $p_{s,r}$, evaluated at the time when the signal impinges on LISA, define a plane that contains the six $\hat{n}_l$. Without loss of generality, let us then express all geometric quantities in an $(x,y,z)$ coordinate system where the LISA plane lies along $x$ and $y$. We reflect the source position across the LISA plane by setting $\hat{k}_z \rightarrow -\hat{k}_z$, and multiplying $h$ on both sides by $\mathrm{diag}(1,1,-1)$; this has the side-effect of rotating the polarization angle $\psi$ of the source.\footnote{For a suitable definition of the polarization angle (as given in Appendix\ A of \cite{Vallisneri2005}), the rotation is just $\psi \rightarrow -\psi$. Now, a generic non-linearly polarized signal can be described by the linear combination of two orthogonally polarized signals; the effect of the reflection considered here is then not just an overall rotation, but also a relative sign change between the two polarizations. This destroys the reflection degeneracy for generic sources, unless yet another source parameter can be adjusted to reverse the sign change.} Because the $\hat{n}_l$ have no $z$ component, all the dot products that appear in Eq.\ \eqref{eq:responses} are unchanged, except for the retarded $h$ times: but since the spacecraft positions $p_{r,s}$ can be written as a vector in the $(x,y)$ plane plus the position vector of the LISA center, $R = (p_1 + p_2 + p_3)/3$, the overall effect is that $\mathrm{TDI}(t)$ acquires an additional delay of $-2 \hat{k} \cdot R$.

To summarize, a linearly polarized burst from some given direction is almost perfectly mimicked, in the LISA data, by a burst whose incidence direction is reflected across the LISA plane (as determined at the time when the GWs impinge on LISA),  and whose polarization  and arrival-time at the SSB are suitably rotated and time-translated, respectively.
This degeneracy is immediately evident as the reflection symmetry across the equator in all the plots in Fig.\ \ref{fig:NoiselessCase}, which examines the $F$-statistic structure for the strongest source in the noiseless training data set. Even for the full LISA response (without any assumptions of stationarity), the reflection symmetry is accurate to better than one part in $10^6$ (in FF), which means that SNRs $\sim$ 1,000 would be required to discriminate between the two sky positions.

\subsection{Broad $F$-statistic quasi-degeneracy across the sky}
\label{ss:broadsymmetry}

Our searches revealed an additional, approximate degeneracy in the $(\mathcal{A},\psi,t_C)$-maximized overlap (i.e., the {\emph F}-statistic) between linearly polarized burst signals incoming from an arbitrary sky position, and templates spread in broad patterns across the sky. This approximate degeneracy appears even if we use all three noise-uncorrelated TDI observables $A$, $E$, and $T$ (see e.g. \cite{VCT2008}), and it is worse (i.e., more nearly degenerate) for bursts with lower $f_\mathrm{max}$.

\begin{figure}
\flushright
\includegraphics[width=4.1in]{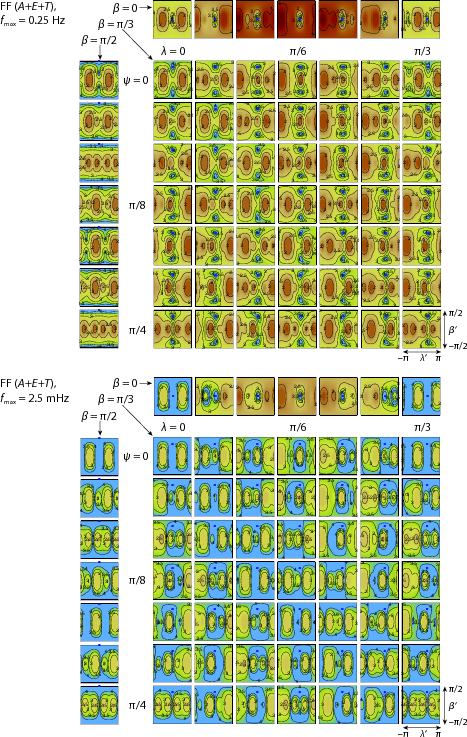}
\caption{FF maps for high- (top) and low-frequency (bottom) bursts: $-\log_{10} (1 - \mathrm{FF})$ contours are computed between $(\beta,\lambda,\psi)$ target sources (with $\beta = 0$, $\pi/3$, $\pi/2$, $\lambda \in [0,\pi/3]$, $\psi \in [0,\pi/4]$) and $(\beta',\lambda')$ templates across the sky ($\beta' \in [-\pi/2,\pi/2]$, $\lambda' \in [-\pi,\pi]$, each small square). Because of the symmetries discussed in Sec.\ \ref{ss:broadsymmetry}, these $\lambda$ and $\psi$ ranges exhaust the variety of maps seen across their entire ranges. The target-source latitude $\beta = \pi/3$ is also representative of latitudes intermediate between the equator $\beta = 0$ and the pole $\beta = \pi/2$. At the equator, $\psi$ has no effect on the maps (except for $\psi = \pi/4$, where there is no LISA response); at the pole, $\lambda$ is degenerate, and $\psi$ is defined consistently with the $\lambda = 0$ meridian. See a zoomable version of this image at \url{http://seadragon.com/view/lmj}.\label{fig:ffmaps}}
\end{figure}

While the reflection degeneracy discussed in the last section has a clear counterpart in the analytical expression of the LISA response to polarized, plane GW waves, this broad degeneracy seems harder to understand analytically. To explore it, in Fig.\ \ref{fig:ffmaps} we present a representative set of \emph{fitting-factor} (FF) sky maps: each map corresponds to a target signal with the sky position and polarization indicated by the dot (and with unit amplitude and arbitrary arrival time); the contours in each map represent the overlap between the target signal and templates across the sky, maximized over the amplitude, polarization, and arrival time of the templates.
By definition, $-1 \leq \mathrm{FF} \leq 1$, but for our signals FF is very close to one across much of the sky, so we actually graph $-\log_{10} (1 - \mathrm{FF})$ (e.g., contour ``4'' corresponds to $\mathrm{FF} = 0.9999$). In all maps (and to label each map) we use latitude and longitude coordinates defined with respect to the instantaneous LISA plane.
To compute the FFs, we work with the frequency-domain representation of burst waveforms \emph{and} of the LISA response, modeling the LISA formation as a stationary, equilateral triangle; this is the same approximation was used in \cite{VCT2008} to compute LISA sensitivity curves. (Unequal armlengths will change the FFs somewhat, but our maps are roughly consistent with the probability distributions found in our searches, which used a full model of the LISA orbits.)  

Looking at Fig.\ \ref{fig:ffmaps}, and specifically at the large square multiple plot at the top (corresponding to a target source with latitude $\beta = \pi/3$), we observe a high-FF cell around the true position of the target source (the dot), with a mirror cell reflected across the LISA plane, at $\beta' = -\beta$.  The two cells sit on a ``circle in the sky'' of higher FF; unlike the case of two ground-based interferometric detectors, this pattern cannot be explained by simple timing considerations, but originates from a more complicated matching of geometric elements. One side of the circle crosses the equator with higher FF, and indeed our searches often yield broken-circle distributions. In the limit of the target source moving to the equator, the two cells coalesce into one; for a target source at the pole, the maps exhibit symmetries that oscillate between two- and four-fold as a function of polarization.
The bottom panel shows that FFs are considerably closer to one for bursts with lower-frequency cut offs, although the structure of the maps is qualitatively the same. 
The appearance of double linked circles in some maps is due to the fact that the highest displayed FF contour is set at 0.9999 (indexed by ``4''); single circles would be seen to form at even higher FF.

We note that Figure \ref{fig:ffmaps} presents sky maps for reduced ranges of the target source's $\lambda$ and $\psi$, which are however representative of the full ranges. Because of a number of symmetries, the map for any $\beta$ and $\lambda$ can be obtained by appropriately shifting and reflecting one of the maps in the figure. To wit (and as exemplified in Fig.\ \ref{fig:showsymmetry}):
\begin{figure}
\flushright\includegraphics[width=0.80\textwidth]{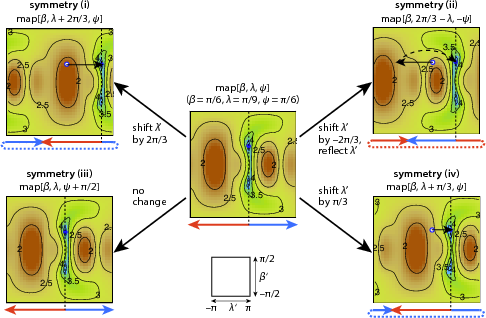}
\caption{Symmetries between FF maps, as explained in the main text, exemplified for the case of $\beta = \pi/6, \lambda = \pi/9, \psi = \pi/6$.\label{fig:showsymmetry}}
\end{figure}
\begin{enumerate}
\item Rotating the source's sky position by $2\pi/3$ around an axis perpendicular to the LISA plane is equivalent to relabeling the three LISA spacecraft (and the TDI observables), so the available geometric information about incoming GW signals must remain the same. Therefore $\mathrm{map}[\beta,\lambda + 2\pi/3,\psi]$ can be obtained by shifting $\mathrm{map}[\beta,\lambda,\psi]$ circularly by $2\pi/3$ along $\lambda'$. This degeneracy was first mentioned in \cite{KeyCornish2008}.
\item Furthermore, there is symmetry in the geometric relation between the LISA spacecraft and sources on either side of a LISA triangle bisector: \includegraphics[height=18pt]{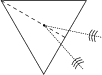}. With the definition of polarization given in \cite{Vallisneri2005}, this results in $\mathrm{map}[\beta,\lambda,\psi]$ reproducing $\mathrm{map}[\beta,2\pi/3 - \lambda,-\psi]$, modulo a $\lambda'$ reflection and circular shift by $2\pi/3$. 
\item Moving on to polarization, letting $\psi \rightarrow \psi + \pi/2$ amounts to reversing the sign of the polarization tensor, a change that is absorbed by the {\emph F}-statistic. It follows that $\mathrm{map}[\beta,\lambda,\psi+\pi/2] = \mathrm{map}[\beta,\lambda,\psi]$.
\item Last, there is a non-obvious symmetry corresponding to reversing the sign of $k$ and $\psi$ for both target source and templates (i.e., to considering signals incoming from the antipodal sky position). Because the burst GWs are invariant w.r.t.\ time inversion about $t_C$, it turns out that the LISA response to $(-k,-\psi)$ signals equals the time-inverted and time-shifted response to the original $(k,\psi)$ signals (see the Appendix). Now, the inner product \eqref{inner} is manifestly invariant w.r.t.\ the time inversion and translation of both $u$ and $v$! Thus, this results in $\mathrm{map}[\beta,\lambda + \pi/3,\psi]$ reproducing $\mathrm{map}[\beta,\lambda,\psi]$, modulo a circular shift by $\pi/3$.
\end{enumerate}
Perhaps the most concise way to characterize the breadth of the degeneracy pattern is to plot, for each map, the fraction of the sky with FF below a given level. We do this in Fig.\ \ref{fig:FFhistogram}, where each of the superimposed lines corresponds to a choice of $\lambda$ and $\psi$ across their entire ranges; the target source latitude is kept fixed to the representative value of $\pi/3$. We can see that for high-frequency bursts (left plot), roughly half of the sky has $\mathrm{FF} > 0.995$, and 2\% (about 800 square degrees) has $\mathrm{FF} > 0.9999$. The plot is even more dramatic for low-frequency bursts, where around 25\% has $\mathrm{FF} > 0.9999$.
The significance of high FFs with respect to the determination of the source's sky position is roughly as follows: for the likelihood of any sky position to decrease by a factor $e$, FF must descend below $1 - 1/\mathrm{SNR}_\mathrm{opt}^2$, where $\mathrm{SNR}_\mathrm{opt}$ is the optimal SNR for a given source. Thus $\mathrm{FF} > 0.9999$ contains the relevant uncertainty region for $\mathrm{SNR} \sim 100$.
\begin{figure}
\flushright
\centerline{\includegraphics[width=0.62\textwidth]{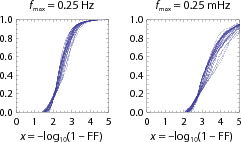}}
\caption{Fraction of the sky with $\mathrm{FF}(A+E+T) > 1 - 10^{-x}$, for target-source $\beta = \pi/3$, and uniformly distributed ($\lambda$,$\psi$), where each pair corresponds to one of the superimposed curves. The curves were obtained by generating $40 \times 40$ maps as for Fig.\ \ref{fig:ffmaps}, assigning a weight to each pixel corresponding to its area in the sky, sorting the resulting sequence by increasing FF, and computing normalized cumulative weights.\label{fig:FFhistogram}}
\end{figure}

\subsection{Effects of degeneracies on searches}
\label{ss:degeneratesearch}

\begin{figure}
\includegraphics[width=0.475\textwidth]{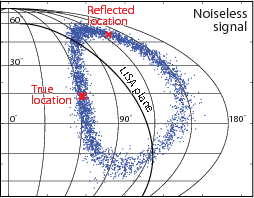}
\hspace{0.04\textwidth}
\includegraphics[width=0.475\textwidth]{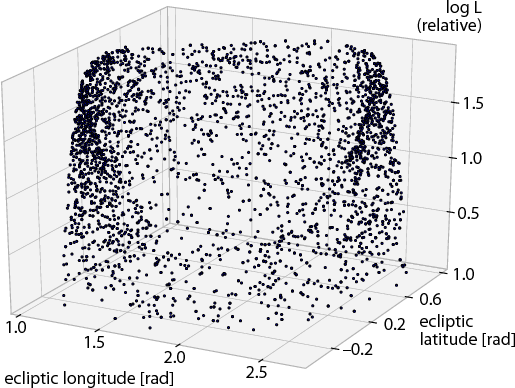}
\caption{Posterior-probability structure for the strongest source (\#3) in the noiseless training data set from MLDC 3.4.
\textbf{Left}: in this sky map, the density of dots (MultiNest equal-weight ``resamples'') is proportional to the posterior probability, maximized over $\mathcal{A}$, $\psi$ and $t_C$, and marginalized over $f_\mathrm{max}$. Crosses mark the true location of the source, and its LISA-plane--reflected counterpart. The map is plotted in the area-preserving Mollweide projection, which we adopt throughout the rest of this paper. 
\textbf{Right}: {\emph F}-statistic as a function of ecliptic latitude and longitude, for the same sky locations as in the left panel. Here {\emph F} is offset by a constant $\simeq 3,029$, and it is only slightly higher for the neighborhoods of the true and reflected sky locations than for the arcs connecting them.
\label{fig:NoiselessCase}}
\end{figure}

The broad quasi-degeneracy pattern is observed clearly in the posterior probability plots produced by our MultiNest runs (see Sec.\ \ref{ss:multinest}). Figure \ref{fig:NoiselessCase} was obtained for the strongest source (with an SNR $\simeq 78$) in the noiseless\footnote{In a truly noiseless data set, the source SNR would be infinite, and it would be possible to determine its source parameters exactly. Figure \ref{fig:NoiselessCase} is instead produced with the usual statistical characterization of noise, for a noise realization that just \emph{happens} to be identically zero.} MLDC 3.4 training data set. In the left-panel sky map, the density of the dots is proportional to the posterior, maximized over $\mathcal{A}$, $\psi$ and $t_C$, and marginalized over $f_\mathrm{max}$. As expected, the dots cluster around the true and reflected locations, but they extend around a thick circle that cuts through the instantaneous LISA plane at the time of the burst. In the right panel, we see that the {\emph F}-statistic decreases only slightly across the circle.

\begin{figure}
\includegraphics[width=\textwidth]{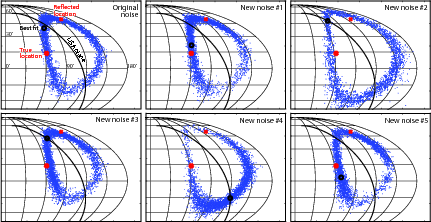}
\caption{Effect of different noise realizations on the posterior-probability structure for the strongest source (\#3) in the noisy MLDC 3.4 training data set, and in five more data sets with the same source and different noise realizations. The additional data sets were created using \texttt{lisatools} \cite{lisatools} with the MLDC 3.4 noise priors, but different pseudorandom-number seeds.
\label{fig:NoiseRealizations}}
\end{figure}

Of course, detector noise will somewhat modify the noiseless posterior distribution. Figure \ref{fig:NoiseRealizations} shows the posteriors computed for the noisy MLDC 3.4 training data set, and for five more data sets with the same source and different noise realizations. Because FFs are consistently high across the circle, it is possible for detector noise to displace the best-fit sky location by large angular distances, while significantly altering the structure of the circle.

\begin{figure}
\includegraphics[width=0.475\textwidth]{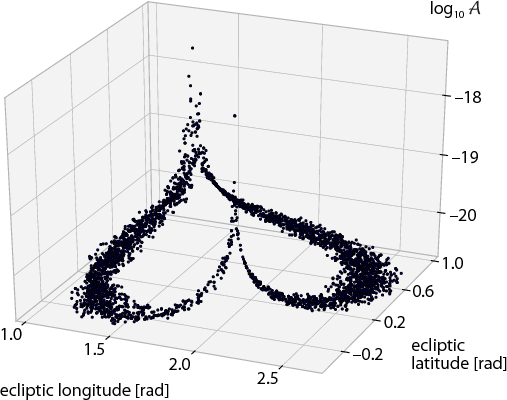}
\hspace{0.04\textwidth}
\includegraphics[width=0.475\textwidth]{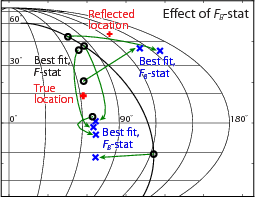}
\caption{\textbf{Left}: the best-fit value for the template amplitude, as computed by the {\emph F}-statistic, increases dramatically for sky positions close to the instantaneous LISA plane, as shown here for source \#3 in the noisy MLDC 3.4 training data set.  \textbf{Right}: the Bayesian $F_B$-statistic shifts the best-fit sky locations away from the instantaneous LISA plane, as seen here for the six data sets of Fig.\ \ref{fig:NoiseRealizations}. In some cases, the best-fit location moves to the other side of the sky; this is not significant, given that reflected points have essentially the same posterior probability against the same source.
\label{fig:EffectOfPrior}}
\end{figure}

In three of the plots of Fig.\ \ref{fig:NoiseRealizations}, the best-fit point ends up very close to the instantaneous LISA plane. Now, sources from those locations elicit a strongly suppressed response in the TDI observables, because they come close to being cross-polarized with respect to the LISA arms. However, by construction the {\emph F}-statistic will raise the template amplitude correspondingly to achieve a good fit to the signal, as shown in the left panel of Fig.\ \ref{fig:EffectOfPrior} for the strongest source (\#3) in the (noisy) MLDC 3.4 training data set. Thus, a ``straight'' maximum-likelihood search can easily lead to a best-fit $\mathcal{A}$ that is orders of magnitude larger than its true value. We have dubbed this phenomenon a \emph{mirage}, because it makes sources appear much stronger and closer than they truly are.

It seems that mirages were not noticed by the other research groups who participated in the MLDC 3 searches for string-cusp bursts~\cite{KeyCornish2008,Feroz_strings_2009}. We conjecture that the reason is as follows. While the {\emph F}-statistic provides the best-fit $\mathcal{A}$ and $\psi$ for any sky location and $f_\mathrm{max}$, the other groups used stochastic algorithms that treat all parameters alike.
Since the mirage occurs in regions of parameter space that are far removed from the true parameters, and in a subspace in which the $\mathcal{A}$ and $\psi$ parameters are rather precisely correlated, it is difficult for these searches to end up in these regions. (Given sufficient time, they \emph{would} arrive there, but if one did not know that the mirages existed, one could easily be fooled into thinking that the search had converged before it actually had.)

Such mirages motivated our development of the Bayesian $F_B$-statistic (Sec.\ \ref{ss:BayesianFstat}), which penalizes the large-amplitude, nearby-source fits that are a priori very unlikely. Best-fit sky locations are correspondingly pushed away from the instantaneous LISA plane, as illustrated in the right panel of Fig.\ \ref{fig:EffectOfPrior} for the six signal-cum-noise realizations of Fig.\ \ref{fig:NoiseRealizations}. Unfortunately, while $F_B$ does tend to disfavor mirage-like fits, it does not necessarily lead to best fits that are any closer to the true locations. The broad quasi-degeneracy described in Sec.\ \ref{ss:broadsymmetry} implies that good fits exist over much of sky, even when Bayesian priors are called into play.

\section{Search methods}
\label{s:Methods}
\subsection{Markov Chain Monte Carlo}
\label{ss:MCMC}
Markov Chain Monte Carlo (MCMC) methods are used to efficiently integrate (and by extension, explore) arbitrary functions $f$ defined over moderate-to-large--dimensional spaces with complex or computationally expensive integration measures $P$ \cite{Liu2001}, when neither analytic techniques nor simple gridding techniques are feasible.
MCMC methods work by creating a \emph{Markov chain} of points that are asymptotically distributed according to $P$. Each next point in the chain is chosen by proposing a new candidate randomly as a function of the current point, and by choosing either the current point or the candidate on the basis of an appropriate criterion that involves their $P$.
For any function $f$ with finite expectation value with respect to $P$ and for sufficiently long chains, the average value of $f$ on the chain approaches the $P$-weighted average of $f$ on the full space.

In applications of MCMC methods to Bayesian inference in signal analysis \cite{ChristensenMeyer1998}, $P$ is typically the posterior probability. In this paper, $P$ is either $e^F$ or $e^{F_B}$, evaluated on the
3-dimensional parameter space $(\beta,\lambda,f_{max})$, or sometimes a subspace. Our Metropolis--Hastings MCMC searches were performed using the \emph{PyMC} software package \cite{PyMC} for the Python programming language. We computed $F$ and $F_B$ as described in Sec.\ \ref{s:theory}, using \emph{Synthetic LISA} \cite{Vallisneri2005} to obtain the GW polarizations $\textbf{h}_{1,2}(t_C)$.
\textit{Synthetic LISA} was designed to perform highly accurate calculations of LISA's TDI responses for any gravitational waveform impinging on LISA (e.g., for burst waveforms it does not use the approximation that LISA is stationary over the timescale of the burst), but this generality and accuracy come at some cost in speed; we find that each computation of $F(t_C)$ or $F_B(t_C)$ takes 2--3 seconds on a $\simeq 3$ GHz processor.
Since single MCMC chains cannot be easily parallelized, we typically compute multiple chains, with each chain beginning in a different location in the parameter space.  

Given a data set, we find it useful to initially localize the bursts in time, at least roughly.  To do this, we create a waveform template with arbitrary values for the sky position $(\beta,\lambda)$ and $f_{\rm max}$, and compute $F(t)$ for all possible times $t$ using the standard inverse Fourier transform trick described in Sec.~\ref{ss:Maximization}.
The peaks of $F(t)$ correspond to the best matches for the template in the data set.  In a search on actual LISA data, we would need to carefully choose a detection threshold, to separate true GW bursts from random noise peaks.  However, because MLDC 3.4 was the first challenge involving a search for cosmic strings in Mock LISA Data, the SNRs of the injected bursts were sufficiently high that the peaks from the bursts could be found in $F(t)$ by eye. Because the sky-position for our template was arbitrary, the true values of $t_C$ (the arrival times of the signal at the SSB, not at LISA) could differ from the times $t_\mathrm{max}$ that maximize $F(t)$ by up to $\sim 10^3$ s.   In practice, we narrowed the search to time windows  $t_C \in [t_\mathrm{max}-2000\,\mathrm{s},t_\mathrm{max}+2000\,\mathrm{s}]$, using a longer-than-necessary window for additional safety.
We use each $t_\mathrm{max}$ as the starting point for a three-stage search: 
\begin{enumerate}
\item For the first stage, we use the fact that the best-fit value of $f_{\rm max}$ has only very weak dependence on the sky position $(\beta,\lambda)$, so we choose a random sky position and perform a 1-D search over $f_{\rm max}$.  Now, not all the MLDC 3.4 sources have a well-defined $f_{\rm max}$, which is chosen randomly (with uniformly distributed logarithm) between $10^{-3}$ and $10$ Hz. (We noted in Sec.~\ref{ss:fmaxdistribution} that the true prior must scale as $f^{-5/3}_{\rm max}$, but rigorous verisimilitude was not a goal of this Challenge.) Thus, $f_{\rm max}$ can be above the 0.5 Hz Nyquist frequency of the data set, in which case $f_{\rm max}$ cannot be determined, other than to say that is $> 0.5$ Hz.
For those signals with $f_{\rm max}$ below Nyquist, we find that $\sim$ 1,000 iterations are sufficient to
obtain a very good estimate. 
\item We now fix $f_{\rm max}$ to this value, and search over the sky position $(\beta,\lambda)$.  For this second stage, we use eight chains of $\sim$ 1,000 iterations each, starting from different sky locations.  
Because of the reflection symmetry across the LISA plane for burst sources (see Sec.\ \ref{ss:SkyPositionReflectionSymmetry}), two nearly equal local modes are found at this stage. For each mode, the point of highest probability among all chains is chosen as the starting point for the third stage of the search.
\item In this final stage, we search over all three $(\beta,\lambda,f_{\rm max})$, restricting the MCMC proposal distribution to a very narrow Gaussian in order to explore only the immediate vicinity of the starting points. We generate one chain for each of the two modes, and define our best fit as the highest-probability point of both chains.
\end{enumerate}
We note that because of the computational limitations discussed above, none of our MCMC runs performed enough iterations to enter the regime of convergence. Therefore, we regard the chains as searches (maximizations) rather than explorations (integrations), and use the maxima attained by the chains as estimates of the true mode of the distributions.

\subsection{MultiNest}
\label{ss:multinest}
MultiNest~\cite{FerozHobson2007,FerozHobson2008} is a publicly available implementation of the \emph{nested-sampling} algorithm for computing the Bayesian evidence of a model given a set of data. 
Nested sampling works by picking a set of $N$ ``live'' points (typically 1,000) at random from parameter space and then systematically replacing the point with the least $P$ with a randomly chosen point\footnote{This random choice must take into account the prior distributions of the parameters. Indeed, MultiNest requires that the $n$-dimensional parameter space first be mapped into the $n$-dimensional unit hypercube, from which MultiNest draws samples assuming a uniform distribution.  Any non-uniform priors must be taken into account in this mapping.} of higher $P$.
In this way the set of live points is gradually attracted toward the modes of the distribution.  As the algorithm proceeds, the number of random draws required to find a suitable replacement for the worst point tends to increase sharply.  In order to alleviate this problem, MultiNest groups live points into ellipses, using the $k$- and $x$-means point-clustering algorithms \cite{Hartigan1975}.  The ellipses are designed to identify and encompass the regions of parameter space that will attract a high concentration of live points. The proposed replacements are then drawn randomly not from the entire space, but from these ellipses.

Nested sampling, like MCMC, provides a way to converge efficiently onto the (local) modes of a distribution.  While this method was designed primarily to calculate the Bayesian evidence (an important concern to determine detection confidence for weak sources), we find that it also performs well at locating local maxima.
Indeed, we found it relatively simple to implement a MultiNest-based search for cosmic-string bursts.  Again, since we use the {\emph F}-statistic and the FFT trick to maximize the likelihood over $(\mathcal{A}, \psi, t_C)$, we define $P$ as $e^F$ or $e^{F_B}$, and search on the remaining three parameters $(\beta,\lambda,f_{\rm max})$.
With 1,000 live points, we find that the code converges well after approximately 10,000 point replacements, or 10 replacements per live point.

Since the probability function is identical to that used for our PyMC searches, the results from the two methods should be in good agreement.  We found that this was indeed the case for both the training and challenge data. 
However, we prefer our MultiNest-based search, for several reasons.  First, it is easily parallelized. While multiple CPUs can be used for multiple chains in MCMC, the long computation time for the log-likelihood results in none of our chains reaching the convergent regime in a reasonable run time.  Although techniques such as parallel tempering and chain mixing increase the utility of a multi-chain approach, they require significantly longer chains than we were able to achieve given our choice to use exact templates (as computed with \emph{Synthetic LISA}) rather than their static-LISA approximation. By comparison, we can easily leverage multiple CPUs for significant speed gains in MultiNest, where multiple candidate replacement points can be prepared in parallel, and unexamined candidates saved for later use.
Second, since our MCMC chains do not reach the convergent regime (as discussed in Sec.~\ref{ss:MCMC}), we are more confident in the results provided by the MultiNest algorithm, which does converge according to a well-defined criterion (a tolerance on the computed evidence).  Finally, MultiNest performs well even without the somewhat elaborate three-stage procedure we use with PyMC.

\subsection{High-SNR limit and the Fisher-Matrix formalism}
\begin{figure}
\centerline{\includegraphics[width=0.50\textwidth]{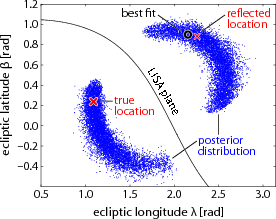}} \vspace{0.2cm}
\includegraphics[width=\textwidth]{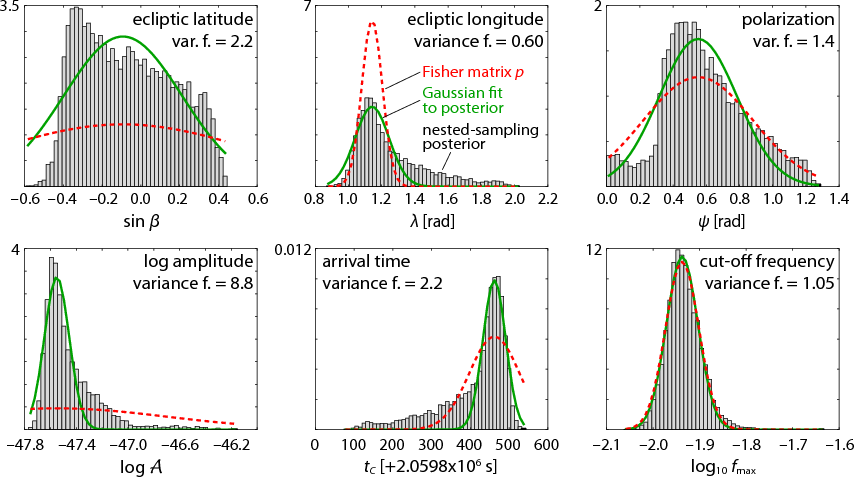}
\caption{Comparison of MultiNest posterior distributions with Fisher-matrix estimates, in the case of the strongest source ($\# 3$) of the MLDC 3.4 noiseless training data.
The top plot shows that the posterior distribution on the sky is more ``banana-shaped'' than ellipsoidal.  The next six plots compare the true posterior distribution (restricted to the neighborhood of the ``true'' mode) with Gaussian distributions of variance $\sigma^2_{\mu} = \big(\Gamma^{-1}\big)^{\mu\mu}$. 
The variance factor, defined as $\sigma^2_{\rm fit} / \sigma^2_{\rm Fisher}$, ranges between $0.6$ and $8.8$.
\label{fig:Fisher_SixD}}
\end{figure}

\begin{figure}
\centerline{\includegraphics[width=0.33\textwidth]{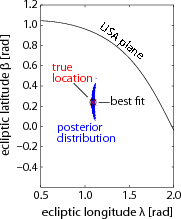}} \vspace{0.2cm}
\includegraphics[width=\textwidth]{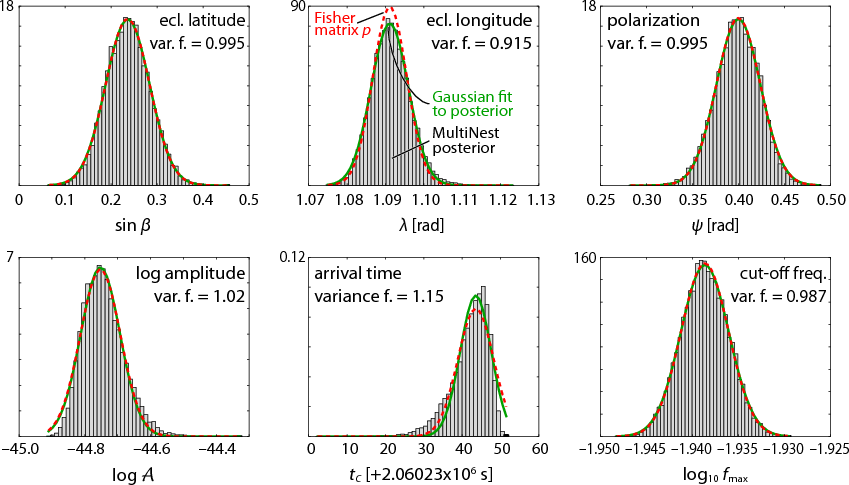}
\caption{
Same as Fig.\ \ref{fig:Fisher_SixD}, except that the source's SNR is now 1,000. In this case, the posterior is fit very well by the Fisher-matrix prediction. Even at this high SNR, a secondary maximum is present around the reflected location, but it is not shown in this plot.
\label{fig:Fisher_SNR1000_SixD}}
\end{figure}

For signals with sufficiently high SNR, the Fisher-matrix formalism provides a useful test of how accurately our codes are calculating the posterior probability.  Consider a single burst immersed in noise, and imagine dialing up the burst's amplitude.  As the SNR increases, the contour of constant likelihood that encloses a given fraction of the total probability (say, 68\% for the 1-$\sigma$ contour) shrinks to encompass an ever smaller region of parameter space.
(Actually, because of the discrete symmetry described in Sec.~\ref{s:Symmetries},  in our case two disjoint contours shrink onto two distinct regions: one region that is close to the true parameter values, and another that is related to it by reflection across the LISA plane.) The smaller the region, the better the log-likelihood function within the contour is described by a constant (the maximum value) plus the second partial derivative term (the Hessian) in a Taylor expansion.  The matrix of partial second derivatives of the log-likelihood is given by
\be\label{second-deriv}
 -\frac{1}{2} \partial_{\mu}\partial_{\nu} \bigl\langle \mathbf{s} - \mathbf{h} \,\big|\, \mathbf{s} - \mathbf{h} \bigr\rangle   \, = \, \left\langle \partial_{\mu}\partial_{\nu} \mathbf{h} \,|\, \mathbf{s} - \mathbf{h} \right\rangle - \Gamma_{\mu\nu} \, ,
\ee
\noindent where $\Gamma_{\mu\nu}$ is the Fisher matrix \cite{useabuse}, defined by  
\begin{equation}
\Gamma_{\mu\nu}  \equiv  \Bigl\langle \frac{\partial}{\partial x^{\mu}} {\bf h} \Big| \frac{\partial}{\partial x^{\nu}} {\bf h} \Bigr\rangle \, .
\end{equation}
Here $\textbf{h}(x^\mu)$ is the waveform (a function of all the parameters $x^\mu$), ${\left\langle \cdots  \,|\, \cdots  \right\rangle}$ is the inner product defined in Eq.~(\ref{inner}), and the partial derivatives are evaluated at the local maximum.
[In a slight abuse of notation, we are using Greek indices to distinguish the Gamma matrix $\Gamma_{\mu\nu}$ on the full parameter space from its restriction to the two-dimensional subspace $(A^1, A^2)$,  which we defined as  $\Gamma_{ij}$ in Sec.~\ref{ss:Maximization}.]
In the high-SNR limit, the posterior distribution function near a local mode approaches a Gaussian, and the second term on the right-hand side of Eq.~(\ref{second-deriv}) dominates, so by integration of a Gaussian exponential the covariance matrix of the parameters (restricted to parameter values near the given mode) approaches the inverse of the Fisher matrix.  To wit: let $x_b^{\mu}$ be the local best-fit parameter values, let  $\Delta x^{\mu} \equiv x^{\mu} - x^{\mu}_b$, and let $\overline{\Delta x^{\mu}\,  \Delta x^{\nu}}$  be the posterior-weighted average of  $\Delta x^{\mu}\,  \Delta x^{\nu}$ (where the averaging is restricted to a neighborhood of the given mode); then
\be
\overline{\Delta x^{\mu}\,  \Delta x^{\nu}} \rightarrow \bigl(\Gamma^{-1}\bigr)^{\mu\nu} \quad
\mathrm{as} \quad \mathrm{SNR} \rightarrow \infty.
\ee
Thus, an especially simple test of the posterior distribution generated by our MultiNest runs is just to check that, for high SNR, the ``variance factor'' $\overline{\Delta x^{\mu} \, \Delta x^{\mu}}/\bigl(\Gamma^{-1}\bigr)^{\mu\mu}$  approaches one for all $\mu$.
As our test case, we choose the strong source (\#3) from the MLDC 3.4 noiseless training data set.  As shown in the top plot in Fig.~\ref{fig:Fisher_SixD}, near both modes the posterior distribution is more ``banana-shaped'' than ellipsoidal, so we would not expect the Fisher-matrix approximation to be very accurate. The bottom six plots in Fig.~\ref{fig:Fisher_SixD} show the posterior distribution for each parameter separately, and compare these with Gaussian distributions based on the inverse Fisher matrix.  We see that in this case, for which the SNR is $\approx 78$, the marginalized posteriors do not have Gaussian shapes, and the Fisher matrix provides only a rough estimate of the actual variances; the variance factor ranges between $0.6$ and $8.8$.
In Fig.~\ref{fig:Fisher_SNR1000_SixD} we show the posterior distribution for the same source, with an increased SNR $\approx$ 1,000. The agreement is much better.

We regard Fig.~\ref{fig:Fisher_SNR1000_SixD} as additional confirmation that our search codes are working as expected. By contrast, we regard Fig.~\ref{fig:Fisher_SixD} as a warning that for LISA detections of string-bursts, even at SNR $\sim 80$, the Fisher-matrix approximation cannot be relied on to predict parameter-estimation errors accurately.

\section{Results from the Mock LISA Data Challenges}
\label{s:Results}

The purpose of the MLDCs is stimulate the development and evaluate the performance of LISA data-analysis tools and methods. In each challenge, data sets containing simulated noise plus GW signals of undisclosed source parameters are made publicly available and all interested research groups are invited to test their algorithms on these \emph{blind} challenge data. Each challenge includes also training data sets with published source parameters, to help groups develop and calibrate their codes. The MLDCs are becoming more realistic with each new challenge, encompassing a larger number and variety of sources.

The third MLDC was the first to include a search for bursts from cosmic strings, MLDC 3.4. This data set consisted of $2^{21}$ samples with a cadence of 1 s (for a total of $\sim 1$ month), and it included a few
randomly chosen string-burst signals injected into purely instrumental noise  (i.e., the data set did \emph{not} include signals from other types of sources, or the confusion noise from unresolvable Galactic binaries). The sky positions of the injected sources were chosen randomly from a uniform sky distribution; the polarizations $\psi$ were drawn uniformly from $[0,\pi]$; and $f_{\rm max}$ was drawn $[10^{-3},\, 10]$ Hz with a uniformly distributed logarithm.

MLDC 3.4 called for a random number (a Poisson deviate of mean $5$) of  injected bursts, with SNRs drawn uniformly from $[10,100]$.  As discussed in Sec.\ \ref{ss:BayesianFstat}, these priors for $f_{\rm max}$ and SNR are not astrophysically realistic, but the intent for this challenge was less to maintain astrophysical realism than to test search algorithms for a wide range
of source parameters (i.e, a wider range than one would obtain from a handful of detections with realistic parameters).
As it turned out, the MLDC 3.4 data set contained exactly three string bursts, all with SNRs in the range $36\mbox{--}45$.  Of course, the realistic expectation is that most detections
will have SNRs within $50\%$ of the detection threshold, which is likely to be $\sim 6$.
Thus, all the MLDC 3.4 bursts had SNRs a factor $4\mbox{--}5$ higher than will be typical.

In this challenge, the exact spectral densities of instrumental noise were randomized and undisclosed, but they were guaranteed to lie within fairly narrow ranges.  In our searches, we ignored this feature, to little apparent damage, by taking the TDI observables to have the standard MLDC noise spectral densities as assumed in the other MLDC challenges.  Explicit expressions for these $S_A(f)$, $S_E(f)$ and $S_T(f)$ are given in \cite{VCT2008}.

\begin{figure}
\includegraphics[width=\textwidth]{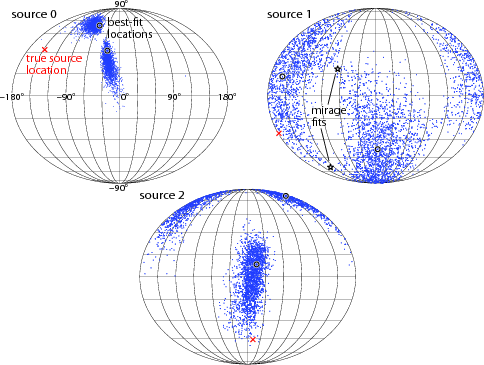}
\caption{MultiNest sky-location posteriors for sources 0--2 in the MLDC 3.4 challenge data set. The density of the dots is proportional to the posterior probability (including the $F_B$ prior correction described in Sec.~\ref{ss:BayesianFstat}), maximized over $t_C$, and marginalized over $\mathcal{A}$, $\psi$, and $f_\mathrm{max}$. Crosses and circles indicate the true and best-fit locations, respectively. For source 1, the stars indicate the location of \emph{mirage} best fits discarded by $F_B$.
\label{fig:Source012_Mollweide}}
\end{figure}

In our entries to MLDC 3.4 and in this paper,\footnote{The values shown in this paper are somewhat different from the values we submitted for MLDC 3.4, which can be viewed at \url{www.tapir.caltech.edu/mldc}.   Our algorithms have improved since the conclusion of MLDC 3, and to keep this paper current with our research effort,  here we have chosen to report our newer results.  In some cases, our newer best-fit parameters are actually further from the true parameter values than our original entries.  Nevertheless, the values reported here arise from a more correct analysis of the data.}
we report the best-fit parameters found by our searches (i.e., the maxima of $F$ or $F_B$). In fact, because there are always two parameter sets that fit the data almost equally well, due to the reflection symmetry described in Sec.~\ref{ss:SkyPositionReflectionSymmetry}, for each burst we report the best-fit parameters of both modes.
Table \ref{tab:BestFitValues} lists the true and best-fit parameters, and Table \ref{tab:ParameterAccuracy} the corresponding estimation errors; Figure \ref{fig:Source012_Mollweide} shows sky plots of the posterior distributions derived from our MultiNest searches.

Certain aspects of the results presented in Table 1 and Figure 10 require clarification. For Source \#0, the true sky location is ruled out by parameter estimation. This should not be surprising: in the high-SNR regime, the variance of $\mathrm{SNR}^2$ over the ensemble of noise realizations is of the order of the number of source parameters; thus the likelihood at the best-fit parameters can exceed the likelihood at the true parameters by large exponential factors.
For Source \#1, we find that the maximum of $F_B$ lies outside the two regions of the sky where the posterior probability is concentrated. In the Table we report instead on the maxima that lie \textit{within} the large, high-probability clusters.  The outlying maximum lies close to the LISA plane, and so it resembles the mirages discussed in Sec.~\ref{ss:degeneratesearch}.  In this case, however, the best-fit
amplitude is only a factor of two higher than the true value, so the Bayesian correction term implicit in $F_B$ does not strongly disfavor it. For Source \#0, MultiNest converged to values of $f_\mathrm{max}$ above the Nyquist frequency, although one of the MCMC chains managed to lock onto a better value.

In summary, we find that both the PyMC and MultiNest searches perform well at locating the peaks of the posterior, and that the best fits found by the two methods are mostly consistent.   In this sense, both techniques are successful. However, because of the broad degeneracy of the posterior across the sky (described in Sec.~\ref{ss:broadsymmetry}), we find that instrument noise will generally shift the best-fit parameters rather far from their true values.  Because the LISA response introduces strong correlations between sky position and the parameters $({\mathcal A}, \psi, t_C$), these come to have
large errors as well. Thus, we should not hope for accurate sky locations in LISA detections of string bursts with SNR $\sim 40$, and the situation will only be worse for typical LISA detections with SNR $\lesssim 10$.

We emphasize that we believe that these large parameter-estimation errors are \emph{not} a result of bugs or lack of convergence in our search methods, but are simply the consequence of the broad parameter-space degeneracy of cusp-burst signals. Besides the consistency between our PyMC and MultiNest results, we performed an additional test by verifying that parameter-estimation accuracy improves when we boost the SNR to $\sim$ 1,000, as shown in Table \ref{tab:TrainingParameterAccuracy} for source $\#3$ in the noisy MLDC 3.4 training data set. For such high SNR, the MultiNest best-fit parameters are reassuringly close to the true values. 

\begin{table}
\caption{\label{tab:BestFitValues} True source parameter values and MCMC and MultiNest best fits for the MLDC 3.4 challenge data set. When the estimated $f_\mathrm{max}$ is larger than the 0.5 Hz Nyquist frequency.
}
\small
\centerline{\begin{tabular}{l@{}r|r@{.}l|r@{.}lr@{.}l|r@{.}lr@{.}l}
\br
\multicolumn{2}{c|}{parameter} & \multicolumn{2}{c|}{true value} & \multicolumn{2}{|c}{MCMC \#1} & \multicolumn{2}{c|}{MCMC \#2} & \multicolumn{2}{|c}{MN \#1} & \multicolumn{2}{c}{MN \#2}\\
\mr
\multicolumn{12}{c}{\textbf{Source 0}} \\
$\beta$       & [rad]                & $0$ & $556$ & $0$ & $551$ & $0$ & $119$ & $0$ & $543$ & $0$ & $933$\\
$\lambda$     & [rad]                & $3$ & $711$ & $5$ & $843$ & $0$ & $005$ & $5$ & $858$ & $5$ & $295$\\
$f_{\rm max}$ & [Hz]                 & $0$ & $030$ & $>0$ & $5$ & $0$ & $044$ & $>0$ & $5$ & $>0$ & $5$\\
$\psi$        & [rad]                & $3$ & $319$ & $2$ & $936$ & $2$ & $776$ & $2$ & $926$ & $1$ & $914$\\
$\mathcal{A}$ & [$10^{-21}$]         & $0$ & $86636$ & $3$ & $0368$ & $1$ & $1394$ & $2$ & $903$ & $3$ & $142$\\
$t_C$         & [$10^6\,\mathrm{s}$] & $1$ & $60216$ & $1$ & $60288$ & $1$ & $60305$ & $1$ & $60289$ & $1$ &$60265$\\
$\rm{SNR}$    &                      & $44$ & $610$ & $44$ & $985$ & $44$ & $842$ & $44$ & $987$ & $44$ & $993$\\
\mr
\multicolumn{12}{c}{\textbf{Source 1}} \\
$\beta$       & [rad]                & $-0$ & $444$ & $-0$ & $753$ & $0$ & $256$ & $-0$ & $658$ & $0$ & $221$\\
$\lambda$     & [rad]                & $3$ & $167$ & $0$ & $015$ & $3$ & $486$ & $0$ & $076$ & $3$ & $502$\\
$f_{\rm max}$ & [Hz]                 & $0$ & $0010842$ & $0$ & $0010927$ & $0$ & $0010932$ & $0$ & $001087$ & $0$ & $001085$\\
$\psi$        & [rad]                & $5$ & $116$ & $4$ & $233$ & $5$ & $023$ & $4$ & $275$ & $5$ & $019$\\
$\mathcal{A}$ & [$10^{-21}$]         & $2$ & $7936$ & $1$ & $6528$ & $1$ & $6585$ & $1$ & $621$ & $1$ & $688$\\
$t_C$         & [$10^6\,\mathrm{s}$] & $1$ & $07269$ & $1$ & $07349$ & $1$ & $07266$ & $1$ & $07352$ & $1$ & $07265$\\
$\rm{SNR}$    &                      & $36$ & $691$ & $36$ & $704$ & $36$ & $702$ & $36$ & $703$ & $36$ & $704$\\
\mr
\multicolumn{12}{c}{\textbf{Source 2}} \\
$\beta$       & [rad]                & $-0$ & $800$ & $0$ & $179$ & $1$ & $154$ & $0$ & $141$ & $1$ & $176$\\
$\lambda$     & [rad]                & $0$ & $217$ & $0$ & $271$ & $2$ & $746$ & $0$ & $259$ & $2$ & $876$\\
$f_{\rm max}$ & [Hz]                 & $6$ & $1495$ & $0$ & $030$ & $0$ & $025$ & $0$ & $026$ & $0$ & $030$\\
$\psi$        & [rad]                & $4$ & $661$ & $4$ & $631$ & $5$ & $225$ & $4$ & $630$ & $5$ & $129$\\
$\mathcal{A}$ & [$10^{-21}$]         & $0$ & $85403$ & $1$ & $0319$ & $1$ & $0285$ & $1$ & $007$ & $1$ & $016$\\
$t_C$         & [$10^6\,\mathrm{s}$] & $0$ & $60001$ & $0$ & $60015$ & $0$ & $59949$ & $0$ & $60015$ & $0$ & $59949$\\
$\rm{SNR}$    &                      & $41$ & $378$ & $41$ & $497$ & $41$ & $496$ & $41$ & $495$ & $41$ & $496$\\
\br
\end{tabular}}
\end{table}

\begin{table}
\caption{\label{tab:ParameterAccuracy} Differences between true
  source parameter values and MCMC and MultiNest best fits, for the MLDC 3.4 challenge data set. The $\Delta \mathrm{sky}$ error is measured in radians along the geodesic arc between the true and best-fit sky positions.}
\small
\centerline{\begin{tabular}{lr|r@{.}lr@{.}l|r@{.}lr@{.}l}
\br
parameter & & \multicolumn{2}{|c}{MCMC \#1} & \multicolumn{2}{c|}{MCMC \#2} & \multicolumn{2}{|c}{MN \#1} & \multicolumn{2}{c}{MN \#2}\\
\mr
\multicolumn{10}{c}{\textbf{Source 0}} \\
$\Delta \mathrm{sky}$     & [rad]  & $1$ & $680$ & $2$ & $278$   & $1$ & $695$ & $1$ & $140$ \\
$\Delta \log_{10} f_{\rm max}$ &   & $>1$ & $222$ & $0$ & $169$   & $>1$ & $222$ & $>1$ & $222$ \\
$\Delta \psi$             & [rad]  & $0$ & $383$ & $0$ & $543$   & $0$ & $394$ & $1$ & $405$ \\
$\Delta \log \mathcal{A}$ &        & $1$ & $254$ & $0$ & $274$   & $1$ & $209$ & $1$ & $288$ \\
$\Delta t_C$              & [s]    & $716$ & $38$ & $881$ & $18$ & $722$ & $40$ & $485$ & $39$ \\
$\Delta \mathrm{SNR}$     &        & $0$ & $375$ & $0$ & $232$   & $0$ & $378$ & $0$ & $383$ \\
\mr
\multicolumn{10}{c}{\textbf{Source 1}}\\
$\Delta \mathrm{sky}$     & [rad]  & $1$ & $944$ & $0$ & $766$                         & $2$ & $039$ & $0$ & $742$ \\
$\Delta \log_{10} f_{\rm max}$ &   & $3$ & $37\times10^{-3}$ & $3$ & $59\times10^{-3}$ & $1$ & $270\times10^{-3}$ & $4$ & $083\times10^{-4}$ \\
$\Delta \psi$             & [rad]  & $0$ & $884$ & $0$ & $093$                         & $0$ & $842$ & $9$ & $758\times10^{-2}$ \\
$\Delta \log \mathcal{A}$ &        & $0$ & $525$ & $0$ & $521$                         & $0$ & $544$ & $0$ & $504$ \\
$\Delta t_C$              & [s]    & $794$ & $28$ & $41$ & $06$                        & $828$ & $39$ & $43$ & $95$ \\
$\Delta \mathrm{SNR}$     &        & 0 & 014 & 0 & 011                                 & $0$ & $012$ & $0$ & $013$ \\
\mr
\multicolumn{10}{c}{\textbf{Source 2}} \\
$\Delta \mathrm{sky}$     & [rad]  & $0$ & $980$ & $2$ & $662$   & $0$ & $942$ & $2$ & $690$ \\
$\Delta \log_{10} f_{\rm max}$ &   & $2$ & $316$ & $2$ & $396$   & $2$ & $377$ & $2$ & $318$ \\
$\Delta \psi$             & [rad]  & $0$ & $030$ & $-0$ & $564$  & $0$ & $031$ & $0$ & $467$ \\
$\Delta \log \mathcal{A}$ &        & $0$ & $189$ & $0$ & $186$   & $0$ & $165$ & $0$ & $174$ \\
$\Delta t_C$              & [s]    & $141$ & $40$ & $519$ & $79$ & $145$ & $06$ & $522$ & $02$ \\
$\Delta \mathrm{SNR}$     &        & 0 & 119 & 0 & 118           & $0$ & $117$ & $0$ & $118$ \\
\br
\end{tabular}}
\end{table}

\begin{table}
\caption{\label{tab:TrainingParameterAccuracy} Parameter accuracy achieved by MultiNest for source \#3 in the MLDC 3.4 training data set, with the original and boosted SNR.}
\small
\flushright\begin{tabular}{l@{}r|r@{.}lr@{.}l|r@{.}lr@{.}l}
\br
\multicolumn{2}{c|}{parameter} & \multicolumn{2}{c}{true value} & \multicolumn{2}{c|}{boosted} & \multicolumn{2}{|c}{best fit} & \multicolumn{2}{c}{best fit (boosted)} \\
\mr
$\beta$       & [rad]                & $0$  & $239$               & \multicolumn{2}{c|}{} & $-0$ & $036$ & $0$ & $247$\\
$\lambda$     & [rad]                & $1$  & $090$               & \multicolumn{2}{c|}{} & $1$ & $204$ & $1$ & $092$\\
$f_{\rm max}$ & [Hz]                 & $1$  & $152\times 10^{-2}$ & \multicolumn{2}{c|}{} & $1$ & $161\times 10^{-2}$ & $1$ & $151\times 10^{-2}$\\
$\psi$        & [rad]                & $0$  & $399$               & \multicolumn{2}{c|}{} & $0$ & $571$ & $0$ & $394$\\
$\mathcal{A}$ & [$10^{-21}$]         & $2$  & $647$ & $37$ & $26$ & $2$ & $204$ & $37$ & $26$\\
$t_C$         & [$10^6\,\mathrm{s}$] & $2$  & $060273$          & \multicolumn{2}{c|}{} & $2$ & $060245$ & $2$ & $060272$\\
$\rm{SNR}$    &                      & $78$ & $122$ & $1082$ & $9278$ & $78$ & $137$ & $1082$ & $9291$\\
\br
\end{tabular}

\begin{tabular}{lr|r@{.}lr@{.}l}
\br
parameter & & \multicolumn{2}{c}{error} & \multicolumn{2}{c}{error (boosted)} \\
\mr
$\Delta \mathrm{sky}$     & [rad]  & $0$ & $297$ & $8$ & $617\times10^{-3}$ \\
$\Delta \log_{10} f_{\rm max}$ &   & $3$ & $6\times10^{-3}$ & $1$ & $2\times10^{-4}$ \\
$\Delta \psi$             & [rad]  & $0$ & $171$ & $4$ & $9\times10^{-3}$ \\
$\Delta \log \mathcal{A}$ &        & $0$ & $183$ & $0$ & $0125$ \\
$\Delta t_C$              & [s]    & $27$ & $89$ & $1$ & $45$ \\
$\Delta \mathrm{SNR}$     &        & $0$ & $015$ & $1$ & $3\times 10^{-3}$ \\
\br
\end{tabular}
\end{table}

\section{Summary, conclusions and future work}
\label{s:Conclusions}
In this paper we have reported on our work to develop two string-burst search pipelines, which rely on the {\emph F}-statistic and the FFT to efficiently maximize the likelihood over $({\mathcal A}, \psi)$ and $t_C$, respectively, and which are based on the publicly available PyMC and MultiNest libraries to maximize over the remaining parameters $(\beta,\lambda,f_{max})$.  Both of our pipelines proved reasonably efficient (MultiNest more so, due to greater gains from parallelization).  We tested our searches by checking that they yielded mutually consistent best fits, and that posteriors results agreed with Fisher-matrix estimates for sufficiently large SNR.   Given the relative simplicity of string-burst signals, we expected that off-the-shelf optimization codes like PyMC and MultiNest be would sufficiently powerful for this search, which our work has verified.

Although the few string-burst injections in MLDC 3.4 had all SNR $\sim 40$, it did not prove possible to localize them on the sky to better than $\sim$ one radian. We showed that this result is just what should be expected, on the basis of the broad degeneracy illustrated by the fitting-factor maps of Sec.~\ref{ss:broadsymmetry}.  Determinations of  ${\mathcal A}$  and $\psi$ are correspondingly poor---again to be expected, since these parameters are strongly correlated with the sky location in the signal measured by LISA. While so far we have analyzed only a handful of bursts in detail, there is every reason to presume that poor parameter-estimation accuracy will be a robust feature of LISA string-burst detections.
In future work, we intend to verify or disprove this presumption by analyzing a much larger sample of bursts drawn from an astrophysically sensible distribution.
This paper also included:
\begin{enumerate}
\item the proof of the near-degeneracy between (linearly polarized) burst signals from directions that are reflections of each other across the LISA plane (which had been noted elsewhere, but heretofore not explained analytically);
\item the first detailed look at string-burst fitting factors as a function of sky position, revealing very high FF over a large fraction of the sky;
\item the analysis of four discrete symmetries (three of which not previously discussed) \emph{between} different fitting function maps;
\item the derivation of an approximate, easily computed Bayesian version of the $F$-statistic, based on realistic priors;
\item a calculation of the expected distribution of $f_{\rm max}$ for detected bursts.
\end{enumerate}

We envisage two broad directions for future work. First, so far we have concentrated on finding the physical parameters of a single string-burst.  Using these sorts of results as input, the next step will be to determine how well LISA can answer questions about the string network (e.g., are there different types of strings? What are $\mu$ and $p$ for each class?) based on an observed population of string-bursts, plus any information from a cosmic-string stochastic background.
Second, so far our searches have been designed for single bursts in Gaussian noise of known spectral density.  We need to generalize our methods to the cases where the noise level and shape are not precisely known (and so must be determined from the data), and where the burst signals are superimposed on a realistic LISA data set containing confusion noise from millions of individually unresolvable sources (mostly white-dwarf binaries) plus tens of thousands of resolvable signals from a variety of sources (especially white-dwarf binaries, EMRIs, and  merging massive black binaries).

\ack This research relied crucially on the hard work of the developers of open-source scientific software: we thank the authors and maintainers the of NumPy and SciPy Python libraries, and especially F.\ Feroz and colleagues (MultiNest) and  C.\ Fonnesbeck and colleagues (PyMC). We thank J.\ Gair, N.\ Cornish, and J.\ Shapiro Key for useful interactions. The numerical computation for this work was performed primarily with computing resources from the Caltech Center for Advanced Computing Research.  MC, CC, and MV all gratefully acknowledge support from NASA grant NNX07AM80G.  Copyright 2009.  All rights reserved.

\appendix

\section{Proof of the fourth FF-map symmetry}

A simple way to see this is to consider a one ``arm'' or a simple-Michelson TDI response
(this entails no loss of generality, since Michelson TDI variables are a basis for all possible observables \cite{VCT2008}, and the derivation would proceed very similarly for first- and second-generation TDI Michelson variables). For instance, using the notation of \cite{Vallisneri2005} and of Eq.\ \eqref{eq:responses}, consider
\begin{eqnarray} \fl
\mathrm{arm}_{12}(k,\psi;t) &=& y_{231}(t) + y_{13'2}(t - L) \nonumber \\
&=& \frac{1}{2} \frac{n_3 \cdot [h(t - k \cdot p_1) - h(t - L - k \cdot p_2)] \cdot n_3}{1 - k \cdot n_{3}} \\
&+& \frac{1}{2} \frac{n_{3'} \cdot [h(t - L - k \cdot p_2) - h(t - 2L - k \cdot p_1)] \cdot n_{3'}}{1 - k \cdot n_{3'}}; \nonumber
\end{eqnarray}
Now $n_{3} = -n_{3'}$, and the dot product of $n_3 \otimes n_3$ with the polarization tensor for a linearly polarized plane GW with $(k,\psi)$ and $(-k,-\psi)$ can be seen to be the same using the formulas of \cite[Appendix\ A]{Vallisneri2005}. Let us then drop those products, and concentrate on the time arguments of the $h$, as well as the geometric projection factors $1 - k \cdot n_j$. Now we let $k \rightarrow -k$, exchange $n_3$ with $-n_{3'}$ in the denominator, and time-advance the whole expression by $2L$:
\begin{equation} \fl
\frac{h(t + 2L + k \cdot p_1) - h(t + L + k \cdot p_2)}{1 - k \cdot n_{3'}} +
\frac{h(t + L + k \cdot p_2) - h(t + k \cdot p_1)}{1 - k \cdot n_{3}};
\end{equation}
after time-inverting the argument of the $h$ (without loss of generality, let $t_C = 0$), we can match the terms one by one with the original expression, yielding, Q.E.D.,
\begin{equation}
\mathrm{arm}_{12}(-k,-\psi;t + 2L) = -\mathrm{arm}_{12}(k,\psi;-t).
\end{equation}

\section*{References}
\bibliographystyle{iopart-num}
\bibliography{References/References_LISA}

\end{document}